\documentclass{aa}
\usepackage{graphicx}

\begin{document}

    \title{Kinematics of young stars
    \thanks{Based on data from the Hipparcos astrometry satellite
            (European Space Agency)}}
    \subtitle{II. Galactic spiral structure}

   \author{D. Fern\'andez\inst{1}
      \and F. Figueras\inst{1}
      \and J. Torra\inst{1,2}
          }

   \offprints{D. Fern\'andez, \\
   \email{dfernand@am.ub.es}}

   \institute{Departament d'Astronomia i Meteorologia, Universitat de
              Barcelona, Av. Diagonal 647, E-08028 Barcelona, Spain
          \and
              IEEC (Institut d'Estudis Espacials de Catalunya), 
              Edif. Nexus-104, Gran Capit\`a 2-4, E-08034 Barcelona, Spain
             }

   \date{Received <date> / Accepted <date>}

   \abstract{
The young star velocity field is analysed by means of a galactic model
which takes into account solar motion, differential galactic rotation and
spiral arm kinematics. We use two samples of Hipparcos data, one
containing O- and B-type stars and another one composed of Cepheid
variable stars. The robustness of our method is tested through careful
kinematic simulations. Our results show a galactic rotation curve with a
classical value of $A$ Oort constant for the O and B star sample
($A^{\mathrm{OB}} =$ 13.7-13.8 km s$^{-1}$ kpc $^{-1}$) and a higher value
for Cepheids ($A^{\mathrm{Cep}} =$ 14.9-16.9 km s$^{-1}$ kpc $^{-1}$,
depending on the cosmic distance scale chosen). The second-order term is
found to be small, compatible with a zero value. The study of the
residuals shows the need for a $K$-term up to a heliocentric distance of 4
kpc, obtaining a value $K = -$(1-3) km s$^{-1}$ kpc$^{-1}$. The results
obtained for the spiral structure from O and B stars and Cepheids show
good agreement. The Sun is located relatively near the minimum of the
spiral perturbation potential ($\psi_\odot =$ 284-20$\degr$) and very near
the corotation circle. The angular rotation velocity of the spiral pattern
was found to be $\Omega_{\mathrm{p}} \approx 30$ km s$^{-1}$ kpc $^{-1}$.
   }

   \maketitle

%
\section{Introduction}

Young stars have been traditionally used as probes of the galactic
structure in the solar neighbourhood. Their luminosity makes them visible
at large distances from the Sun, and their age is not great compared to
the dynamical evolution timescales of our galaxy.

These stars show kinematic characteristics that cannot be explained by
solar motion and differential galactic rotation alone. Small perturbations
in the galactic gravitational potential induce the formation of density
waves that can explain part of the special kinematic features of young
stars in the solar neighbourhood and also the existence of spiral arms in
our galaxy. The first complete mathematical formulation of this theory was
done by Lin and his associates (Lin \& Shu \cite{Lin et al.1}; Lin et al.
\cite{Lin et al.2}).

Lin's theory has several free parameters that can be derived from
observations. Two of the main parameters are the number of spiral arms
($m$) and their pitch angle ($i$). Although the original theory proposed a
2-armed spiral structure with $i = -6\degr$, as early as the mid-70s
Georgelin \& Georgelin (\cite{Georgelin et al.}) found 4 spiral arms with
$i = -12\degr$ from a study of the spatial distribution of HII regions.
This controversy is still not resolved: in a review Vall\'ee
(\cite{Vallee}) concluded that the most suitable value is $m = 4$, whereas
Drimmel (\cite{Drimmel}) found that emission profiles of the galactic
plane in the K band --which traces stellar emission-- are consistent with
a 2-armed pattern, whereas the 240 $\mu$m emission from dust is compatible
with a 4-armed structure. In a recent paper, L\'epine et al. (\cite{Lepine
et al.}) described the spiral structure of our Galaxy in terms of a
superposition of 2- and 4-armed wave harmonics, studying the kinematics of
a sample of Cepheids stars and the $l$-$v$ diagrams of HII regions.

The angular rotation velocity of the spiral pattern $\Omega_\mathrm{p}$ is
another parameter of the galactic spiral structure. It determines the
rotation velocity of the spiral structure as a rigid body. The classical
value proposed by Lin et al. (\cite{Lin et al.2}) is $\Omega_\mathrm{p}
\approx$ 13.5 km s$^{-1}$ kpc$^{-1}$. The angular rotation velocity in the
solar neighbourhood due to differential galactic rotation is $\Omega_\odot
\approx$ 26 km s$^{-1}$ kpc$^{-1}$ (Kerr \& Lynden-Bell \cite{Kerr et
al.}). Thus, the value of $\Omega_\mathrm{p}$ implies that the so-called
corotation circle (the galactocentric radius where $\Omega_{\mathrm{p}} =
\Omega$) is in the outer region of our galaxy ($\varpi_{\mathrm{cor}}
\approx$ 15-20 kpc, depending on the galactic rotation curve assumed).  
Nevertheless, several authors found higher values of $\Omega_\mathrm{p}$,
about 17-29 km s$^{-1}$ kpc$^{-1}$ (Marochnik et al. \cite{Marochnik et
al.}; Cr\'ez\'e \& Mennessier \cite{Creze et al.}; Byl \& Ovenden
\cite{Byl et al.};  Avedisova \cite{Avedisova}; Amaral \& L\'epine
\cite{Amaral et al.};  Mishurov et al. \cite{Mishurov et al.1}; Mishurov
\& Zenina \cite{Mishurov et al.2}; L\'epine et al. \cite{Lepine et al.}).
These values place the Sun near the corotation circle, in a region where
the difference between the galactic rotation velocity and the rotation of
the spiral arms is small. This fact has very important consequences for
the star formation rate in the solar neighbourhood, since the compression
of the interstellar medium due to shock fronts induced by density waves
could be chiefly responsible for this process (Roberts \cite{Roberts2}).

Other parameters of Lin's theory are the amplitudes of induced
perturbation in the velocity (in the antigalactocentric and the galactic
rotation directions) of the stars and gas, and the phase of the spiral
structure at the Sun's position. The interarm distance and the phase of
the spiral structure can be determined from optical and radio observations
(Burton \cite{Burton}; Bok \& Bok \cite{Bok et al.}; Schmidt-Kaler
\cite{Schmidt-Kaler}; Elmegreen \cite{Elmegreen}). The interarm distance
gives us a relation between the number of arms and the pitch angle.

In this paper we obtain the galactic kinematic parameters from two samples
of Hipparcos stars described in Sect. \ref{samples}: one that contains O-
and B-type stars, and another one composed of Cepheid variable stars. In
Sect. \ref{model} we propose a model of our galaxy, a generalization of
that previously used by Comer\'on \& Torra (\cite{Comeron et al.1}). The
authors only applied their model to radial velocities. The accurate
astrometry of the Hipparcos satellite offers a good opportunity to also
use the proper motion data. The resolution of the condition equations,
based on a weighted least squares fit, is explained in Sect. \ref{mat}. An
extensive set of simulations is performed in Sect. \ref{test} in order to
assess the capabilities of the method, that is, to analyze the influence
of the observational errors and biases in the kinematic parameters. We
finally present our results and discussion in Sect. \ref{results} and
\ref{discussion}, respectively.

%
\section{The working samples}\label{samples}

\subsection{Sample of O and B stars}

\begin{figure}
  \resizebox{\hsize}{!}{\includegraphics{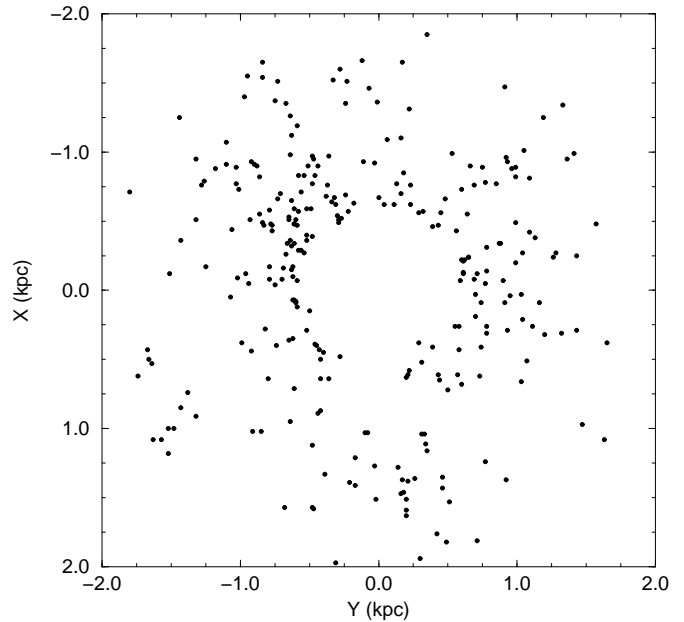}}
  \caption{Star distribution in the $X$-$Y$ galactic plane for the sample
           of O and B stars with 0.6 $< R <$ 2 kpc.}
  \label{fig.xy}
\end{figure}

Our initial sample contains 6922 Hipparcos O- and B-type stars. The
astrometric data for these stars come from the Hipparcos Catalogue (ESA
\cite{ESA2}), radial velocities from the compilation of Grenier
(\cite{Grenier}) and Str\"omgren photometry from Hauck \& Mermilliod's
(\cite{Hauck et al.}) catalogue. The procedure followed to elaborate the
sample considered here is fully described in Torra et al. (\cite{Torra et
al.}; hereafter referred to as Paper I) along with a study of the possible
biases in the trigonometric distances and the availability of radial
velocities. Our sample contains 3915 stars with known distance and proper
motions (2272 stars with known radial velocity). In Paper I we
characterized the structure and kinematics of the Gould Belt system using
this sample of O and B stars, establishing its boundary to a distance of
about 0.6 kpc from the Sun. Taking into account the kinematic
peculiarities of the Gould Belt, in this paper we do not consider those
stars with $R < 0.6$ kpc. In Fig. \ref{fig.xy} we show the position of
those stars with $0.6 < R < 2$ kpc projected on the galactic plane ($X$
positive towards the galactic center and $Y$ positive towards the galactic
rotation direction), which is our working sample (448 stars; 307 of them
with distance, radial velocity and proper motions and 141 with only
distance and proper motions).

\subsection{Sample of Cepheid stars}

The initial sample contains all the Hipparcos classical Cepheids.
Astrometric data were taken from the Hipparcos Catalogue (ESA
\cite{ESA2}), whereas radial velocities come from Pont et al. (\cite{Pont
et al.1}, \cite{Pont et al.2}).

Individual distances were computed following two period-luminosity (PL)
relations. In both, periods come from the Hipparcos Catalogue (ESA
\cite{ESA2}) and individual reddenings from Fernie et al.'s (\cite{Fernie
et al.}) compilation (continuous updating). A classification between
fundamental and overtone Cepheids from light curves and Fourier analysis
was adopted (Beaulieu \cite{Beaulieu}), using only the former in the
kinematic analysis.

The first PL relation (Luri \cite{Luri}) adopts a slope from EROS
(Beaulieu \cite{Beaulieu}) and corresponds to the short cosmic distance
scale:
\begin{eqnarray}
  M_v^{\mathrm{Short}} = -1.08 - 2.72 \log P
\end{eqnarray}

\noindent On the other hand, the second PL relation (Feast \& Catchpole
\cite{Feast et al.0}) corresponds to the large distance scale:
\begin{eqnarray}
  M_v^{\mathrm{Large}} = -1.41 - 3.46 \log P
\end{eqnarray}

\begin{figure}
  \resizebox{\hsize}{!}{\includegraphics{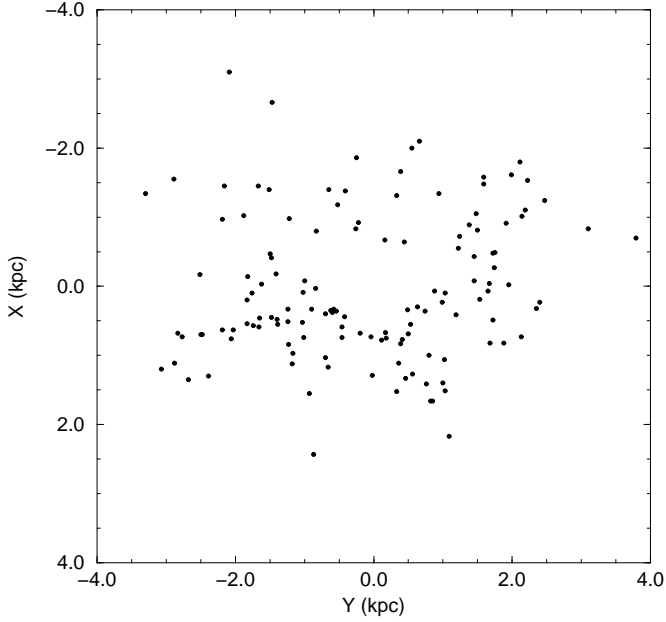}}
  \caption{Star distribution in the $X$-$Y$ galactic plane for the sample
           of Cepheid stars with 0.6 $< R <$ 4 kpc. Distances computed
           from a short cosmic distance scale (Luri \cite{Luri}).}
  \label{fig.xy.cef}
\end{figure}

\noindent The sample contains 186 stars with known distance and proper
motions (165 stars with radial velocity). Their distribution in the
galactic plane (0.6 $< R <$ 4 kpc) is shown in Fig. \ref{fig.xy.cef} (164
stars; 145 of them with distance, radial velocity and proper motions and
19 with only distance and proper motions).

%
\section{A galactic kinematic model for the solar
neighbourhood}\label{model}

\subsection{Systematic velocity field in the galactic model}

We propose an extension of the bidimensional model previously applied for
radial velocities by Comer\'on \& Torra (\cite{Comeron et al.1}). In our
model we considered the systematic velocity components:
\begin{eqnarray}\label{vtot}
  v_{\mathrm{r}} & = & v_{\mathrm{r_1}} + v_{\mathrm{r_2}} + v_{\mathrm{r_3}}
  \nonumber \\
  v_{\mathrm{l}} & = & v_{\mathrm{l_1}} + v_{\mathrm{l_2}} + v_{\mathrm{l_3}}
  \nonumber \\
  v_{\mathrm{b}} & = & v_{\mathrm{b_1}} + v_{\mathrm{b_2}} + v_{\mathrm{b_3}}
\end{eqnarray}

\noindent where subindex 1 refers to the solar motion contribution,
subindex 2 to differential galactic rotation and subindex 3 to spiral arm
kinematics, respectively (see Appendix \ref{ApenA}). Solar motion is
expressed through the three components of the Sun's velocity in galactic
coordinates ($U_\odot, V_\odot, W_\odot$). Galactic rotation curve was
developed up to second-order approximation, $a_\mathrm{r}$ and
$b_\mathrm{r}$ being the first- and second-order terms, respectively:
\begin{eqnarray}\label{eq.ab}
  a_\mathrm{r} & = & \left( \frac {\partial \Theta}{\partial \varpi}
\right)_\odot = \frac {\Theta(\varpi_\odot)}{\varpi_\odot} - 2 A
\nonumber \\
  b_\mathrm{r} & = & \frac {1}{2} \left( \frac {\partial^2
\Theta}{\partial \varpi^2} \right)_\odot
\end{eqnarray}

\noindent where we show the relationship between $a_\mathrm{r}$ and the
$A$ Oort constant. $B$ Oort constant can be derived from:
\begin{equation}
  B = A - \frac {\Theta(\varpi_\odot)}{\varpi_\odot}
\end{equation}

\noindent We considered as free parameters the galactocentric distance of
the Sun ($\varpi_\odot$) and the circular velocity at the Sun's position
($\Theta(\varpi_\odot)$). Finally, spiral arm kinematics was modelled
within the framework of Lin's theory. We considered the number of the
spiral arms ($m$) and their pitch angle ($i$) as free parameters, and we
derived the perturbation velocity amplitudes in the antigalactocentric and
tangential directions ($\Pi_{\mathrm{b}}$ and $\Theta_{\mathrm{b}}$,
respectively; they were considered as constant magnitudes, assuming that
their variation with the galactocentric distance is smooth), the phase of
the spiral structure at the Sun's position ($\psi_\odot$) and the
parameter $f_\odot$, which takes into account the difference in the
velocity dispersion between the solar-type stars and the considered stars.
This parameter is defined by the relationship between the spiral
perturbation velocity amplitudes for the sample stars ($\Pi_{\mathrm{b}}$,
$\Theta_{\mathrm{b}}$) and for the Sun ($\Pi_{\mathrm{b \odot}}$,
$\Theta_{\mathrm{b \odot}}$):
\begin{eqnarray}
   \Pi_{\mathrm{b \odot}} = \frac {1 - \nu^2 +
      x_{\mathrm{stars}}}{1 - \nu^2 + x_\odot} 
      \Pi_{\mathrm{b}} \equiv f_\odot \Pi_{\mathrm{b}} \nonumber \\
   \Theta_{\mathrm{b \odot}} = \frac {1 - \nu^2 +
      x_{\mathrm{stars}}}{1 - \nu^2 + x_\odot}
      \Theta_{\mathrm{b}} \equiv f_\odot \Theta_{\mathrm{b}}
\end{eqnarray}

\noindent where $\nu$ is the dimensionless rotation frequency of the
spiral structure and $x$ is the stability Toomre's number (Toomre
\cite{Toomre}), which depends on the velocity dispersion of the considered
stars (see details in Appendix \ref{ApenA}). The inclusion of $f_\odot$ is
new with regard to the model proposed by Comer\'on \& Torra (\cite{Comeron
et al.1}).

Eqs. (\ref{vtot}) can be expressed as:  
\begin{eqnarray} \label{sistema}
  v_{\mathrm{r}}(R,l,b) & = & \sum\limits_{j=1}^{10}
  a_j f_j^{\mathrm{r}}(R,l,b)
\nonumber \\
  v_{\mathrm{l}}(R,l,b) & = & \sum\limits_{j=1}^{10}
  a_j f_j^{\mathrm{l}}(R,l,b)
\nonumber \\
  v_{\mathrm{b}}(R,l,b) & = & \sum\limits_{j=1}^{10}
  a_j f_j^{\mathrm{b}}(R,l,b)
\end{eqnarray}

\noindent where the constants $a_j$ contain combinations of the kinematic
parameters we wish to determine ($U_\odot$, $V_\odot$, $W_\odot$,
$a_{\mathrm{r}}$, $b_{\mathrm{r}}$, $\psi_\odot$, $\Pi_{\mathrm{b}}$,
$\Theta_{\mathrm{b}}$ and $f_\odot$) and $f_j^i(R,l,b)$ are functions of
the heliocentric distance and the galactic longitude and latitude (see
Eqs. (\ref{eq.a}) and (\ref{eq.f})).

\subsection{Free parameters of our galactic model}
\label{subsect.parameters}

A 2-armed Galaxy was the first proposed view for our stellar system,
mainly derived from HI and HII observations, but also from the spatial
distribution of supergiant stars and other bright objects. These classical
studies show the existence of at least two arms inside the solar circle
(the Sagittarius-Carina or $-$I arm and the Norma-Scutum or $-$II arm),
one local arm (Orion-Cygnus or 0 arm) and one external arm (Perseus or
$+$I arm). The Orion-Cygnus arm seems to be a local spur (Bok \cite{Bok}).
Lin et al. (\cite{Lin et al.2}) proposed a galactic system with 2 main
spiral arms, where the Norma-Scutum and the Perseus arms are two segments
of the same arm. By taking into account the interarm distance between the
Sagittarius-Carina and the Perseus arms, these authors deduced a pitch
angle of $-$6\degr. But, as early as the mid-70s, Georgelin \& Georgelin
(\cite{Georgelin et al.}) proposed a 4-armed galactic system with a pitch
angle of $-12\degr$ from a study of the spatial distribution of HII
regions. However, Bash (\cite{Bash}) examined this 4-armed model and found
that a 2-arm pattern predicts HII regions in the same direction and with
the same radial velocities as those used by Georgelin \& Georgelin
(\cite{Georgelin et al.}), provided that dispersion velocities were
considered.

In some recent papers several authors have also called this classical view
in question. Vall\'ee (\cite{Vallee}) reviewed the subject of the
determination of the pitch angle and the number of spiral arms and
concluded that the Galaxy has a pitch angle of $i = -12 \pm 1\degr$ and
that, taking into account the observed interarm distance, it would be a
system of 4 spiral arms. This is also the opinion expressed by Amaral \&
L\'epine (\cite{Amaral et al.}), who, fitting the galactic rotation curve
to a mass model of the Galaxy, found an autoconsistent solution with a
system of 2 + 4 spiral arms (2 arms for 2.8 $< \varpi <$ 13 kpc and 4 arms
for 6 $< \varpi <$ 11 kpc, with the Sun placed at $\varpi_\odot =$ 7.9
kpc) and a pitch angle of $i = -14\degr$. Englmaier \& Gerhard
(\cite{Englmaier et al.}) used the COBE NIR luminosity distribution and
connected it with the kinematic observations of HI and molecular gas in
$l$-$v$ diagrams. They found a 4-armed spiral pattern between the
corotation of the galactic bar and the solar circle. Drimmel
(\cite{Drimmel}) found that the galactic plane emission in the K band is
consistent with a 2-armed structure, whereas the 240 $\mu$m emission from
dust is compatible with a 4-armed pattern. In a recent work, L\'epine et
al. (\cite{Lepine et al.}) analyzed the kinematics of a sample of Cepheid
stars and found the best fit for a model with a superposition of 2+4
spiral arms. Contrary to the model by Amaral \& L\'epine (\cite{Amaral et
al.}), L\'epine et al. allowed the phase of both spiral patterns to be
independent, deriving pitch angles of approximately $-6\degr$ and
$-12\degr$ for $m=2$ and $m=4$, respectively. They argued that this spiral
pattern is in good agreement with the $l$-$v$ diagrams obtained from
observational HII data, though they admit that pure 2-armed model produces
similar results. In the visible spiral structure of the Galaxy derived by
L\'epine et al. (see their Fig. 3) the Orion-Cygnus or local arm is seen
as a major structure with a small pitch angle. In contrast, Olano
(\cite{Olano}) proposed that the local arm is an elongated structure of
only 4 kpc of length and a pitch angle of about $-40\degr$ (in better
agreement with the observational determinations of the inclination of the
local arm found in the literature) formed from a supercloud about 100 Myr
ago, when it entered into a major spiral arm.

In the case of the galactocentric distance of the Sun and the circular
velocity at the Sun's position there are also some inconsistencies among
the different values found in the literature. In 1986, the IAU adopted the
values $\varpi_\odot =$ 8.5 kpc and $\Theta(\varpi_\odot) =$ 220 km
s$^{-1}$ (Kerr \& Lynden-Bell \cite{Kerr et al.}). Recently, several
authors have found values of nearly 7.5 kpc for $\varpi_\odot$ (Racine \&
Harris \cite{Racine et al.}; Maciel \cite{Maciel}). A complete review was
done by Reid (\cite{Reid}), who concluded that the most suitable value
seems to be $\varpi_\odot = 8.0 \pm 0.5$ kpc. In kinematic studies there
are serious discrepancies between different authors. Metzger et al.  
(\cite{Metzger et al.}) found $\varpi_\odot = 7.7 \pm 0.3$ kpc and
$\Theta(\varpi_\odot) = 237 \pm 12$ km s$^{-1}$, whereas Feast et al.  
(\cite{Feast et al.2}) found $\varpi_\odot = 8.5 \pm 0.3$ kpc (both from
radial velocities of Cepheid stars). Glushkova et al. (\cite{Glushkova et
al.}) found $\varpi_\odot = 7.3 \pm 0.3$ kpc from a combined sample
including open clusters, red supergiants and Cepheids. Olling \&
Merrifield (\cite{Olling et al.}) calculated mass models for the Galaxy
and concluded that a consistent picture only emerges when considering
$\varpi_\odot = 7.1 \pm 0.4$ kpc and $\Theta(\varpi_\odot) = 184 \pm 8$ km
s$^{-1}$. This value for the galactocentric distance of the Sun is in very
good agreement with the only direct distance determination ($\varpi_\odot
= 7.2 \pm 0.7$ kpc), which was made employing proper motions of H$_2$O
masers (Reid \cite{Reid}).

In the view of all that, we decided to derive the kinematic parameters of
our model from different combinations of the free parameters involved. On
the one hand, concerning spiral structure, two models of the Galaxy were
considered: a first model with $m = 2$ and $i = -6\degr$, and a second one
with $m = 4$ and $i = -14\degr$. Both models are consistent with an
interarm distance of about 2.5-3 kpc, depending on the adopted value for
the distance from the Sun to the galactic center. On the other hand,
concerning the galactocentric distance of the Sun and the circular
velocity at the Sun's position, two different cases were also taken into
account: a first one with $\varpi_\odot = 8.5$ kpc and
$\Theta(\varpi_\odot) = 220$ km s$^{-1}$, and a second one with
$\varpi_\odot = 7.1$ kpc and $\Theta(\varpi_\odot) = 184$ km s$^{-1}$. In
both cases, the angular rotation velocity at the Sun's position is
$\Omega_\odot = 25.9$ km s$^{-1}$ kpc$^{-1}$.

%
\section{Resolution of the condition equations}\label{mat}

We determined the kinematic parameters of the galactic model via least
squares fit from the equations:
\begin{eqnarray} \label{eq.fit}
  v_{\mathrm{r}}
  & = & \sum\limits_{j=1}^{10} a_j f_j^{\mathrm{r}}(R,l,b)
\nonumber \\
  v_{\mathrm{l}} = k \, R \, \mu_{\mathrm{l}} \, \cos b 
  & = & \sum\limits_{j=1}^{10} a_j f_j^{\mathrm{l}}(R,l,b)
\nonumber \\
  v_{\mathrm{b}} = k \, R \, \mu_{\mathrm{b}}
  & = & \sum\limits_{j=1}^{10} a_j f_j^{\mathrm{b}}(R,l,b)
\end{eqnarray}

\noindent where $v_{\mathrm{r}}$ is the radial velocity of the star in km
s$^{-1}$, $k$ = 4.741 km yr (s pc $\arcsec$)$^{-1}$, $R$ is the
heliocentric distance of the star in pc, $\mu_{\mathrm{l}}$ and
$\mu_{\mathrm{b}}$ are the proper motion in galactic longitude and
latitude of the star in \arcsec \, yr$^{-1}$ and $b$ the galactic
latitude. To derive the parameters $a_j$, an iterative scheme
extensively explained in Fern\'andez (\cite{Fernandez}) was followed.

The weight system was chosen as:
\begin{equation}
  p_k = \frac {1} {\sigma_{k,\mathrm{obs}}^2 + \sigma_{k,\mathrm{cos}}^2}
\end{equation}

\noindent where $\sigma_{\mathrm{obs}}$ are the individual observational
errors in each velocity component of the star, calculated by taking into
account the correlations between the different variables provided by the
Hipparcos Catalogue, and $\sigma_{\mathrm{cos}}$ is the projection of the
cosmic velocity dispersion ellipsoid in the direction of the velocity
component considered (see Paper I).

To check the quality of the least squares fits we considered the $\chi^2$
statistics for $N - M$ degrees of freeedom, defined as:
\begin{equation}  
  \chi^2 = \sum_{k=1}^{N}
  \frac {\left[ y_k - y(x_k; a_1,...,a_{10}) \right]^2}
  {\sigma_{k,\mathrm{obs}}^2 + \sigma_{k,\mathrm{obs}}^2}
\end{equation}

\noindent where $x_k$ are the independent data (sky coordinates and
distances), $y_k$ the dependent data (radial and tangential velocity
components), $N$ the number of equations and $M$ number of parameters to
be fitted. According to Press et al. (\cite{Press et al.}), if the
uncertainties (cosmic dispersion and observational errors) are
well-estimated, the value of $\chi^2$ for a moderately good fit would be
$\chi^2 \approx N - M$, with an uncertainty of $\sqrt{2 \; (N - M)}$.   

As we did in Paper I, to eliminate the possible outliers present in the
sample due to both the existence of high residual velocity stars (Royer
\cite{Royer}) or stars with unknown large observational errors, we
rejected those equations with a residual larger than 3 times the root mean
square residual of the fit (computed as $ \sqrt{[ y_k - y(x_k; a_1,...,
a_{10})]^2 / N}$) and recomputed a new set of parameters.

%
\section{Test of robustness}\label{test}

The spiral arm potential is expected to contribute between 5-10\% to the
whole galactic gravitational field in the solar neighbourhood.  This small
contribution, and the large observational errors and constraints present
on the spatial and kinematic parameters of distant stars, make it very
difficult to quantify the kinematic perturbations induced by the spiral
potential. Then, the results found in the literature have been
characterised by significant uncertainties and discrepancies. A good
example of this is the contradictory results obtained for the phase of the
spiral structure at the Sun's position (see Sect. \ref{subsect.spiral}):
between arms or near an arm? Interesting questions to answer after the
release of the Hipparcos data are:

\begin{itemize}

\item How does the unprecedented astrometric precision provided by
Hipparcos help to diminish such uncertainties? Are they small enough to
measure the spiral arm kinematic parameters to any useful degree of
accuracy?

\item Can we quantify the biases induced by our observational errors and
constraints?

\item How can the correlations among variables present in the condition
equations affect the kinematic parameters?

\end{itemize}

With regard to our stellar data, as seen in Sect. \ref{samples}, both O
and B star and Cepheid samples suffer from different observational
limitations. Although the O and B star sample is large in number, it is
limited in distance to no more than 1.5-2 kpc from the Sun (see Fig.
\ref{fig.xy} and Table \ref{tab.interval}). In contrast, the Cepheid
sample reaches distances of up to about 4 kpc, but the number of stars
with reliable data still remains very small (Fig.  \ref{fig.xy.cef} and
Table \ref{tab.interval}).

\begin{table}
\caption[]{Number of stars with distance and proper motions (in brackets
           those stars with also radial velocity) in several distance
           intervals. Distances for Cepheids computed from Luri's
           (\cite{Luri}) PL relation.}
\label{tab.interval}
\begin{tabular}{ccc}
	\hline
        \multicolumn{3}{c}{sample of O and B stars} \\
        \hline
        0.1 $< R <$ 2 kpc & 0.6 $< R <$ 2 kpc &                   \\
        \hline
        \hline
        3418 (1903)       & 448 (307)         &                   \\
        \hline
        \hline
        \multicolumn{3}{c}{sample of Cepheid stars} \\
        \hline
        0.1 $< R <$ 4 kpc & 0.6 $< R <$ 2 kpc & 0.6 $< R <$ 4 kpc \\
        \hline
        \hline
        119 (111)         & 103 (95)          & 164 (145)         \\
        \hline
\end{tabular} 
\end{table}

Numerical simulations were performed to answer the above questions, that
is, to evaluate and quantify all the uncertainties and biases involved in
our resolution process. In Appendix \ref{ApenB} we present the detailed
procedure followed to build simulated samples as similar as possible to
the real data for both O and B stars and Cepheids. This Appendix also
contains an exhaustive analysis of the full set of cases which were
simulated. Next we present the main conclusions arising from this work,
which substantially contribute to the analysis of the real data:

\begin{itemize}

\item As a general trend, owing to the correlations between variables, the
biases and uncertainties in some kinematic parameters substantially vary
when changing the real values of $\psi_\odot$ and $\Omega_{\mathrm{p}}$.

\item The first- and second-order terms of the galactic rotation curve are
accurately derived, with an uncertainty of less than 1.3 km s$^{-1}$
kpc$^{-1}$ (km s$^{-1}$ kpc$^{-2}$) in all cases. Whereas a negative bias
for both parameters is detected for O and B stars (up to $-0.7$ km
s$^{-1}$ kpc$^{-1}$ for $a_{\mathrm{r}}$ and $-1$ km s$^{-1}$ kpc$^{-2}$
for $b_{\mathrm{r}}$), it is negligible for Cepheids.

\item For the spiral structure parameters, we obtain better results from O
and B stars than from Cepheids. This is due to the larger number of O and
B stars. For both samples we found a clear dependence of some parameters
with the assumed values for $\psi_\odot$ and $\Omega_{\mathrm{p}}$. We can
expect a bias in $\psi_\odot$ up to $\pm 20\degr$, and uncertainties of
10-20$\degr$ for O and B stars and 30-60$\degr$ for Cepheids. For
$\Pi_{\mathrm{b}}$ and $\Theta_{\mathrm{b}}$ we found biases up to $\pm 2$
km s$^{-1}$ and uncertainties of 1-2 km s$^{-1}$. We did not find an
important bias for $\Omega_{\mathrm{p}}$ neither for O and B stars nor
Cepheids, though its standard deviation is large (4-10 km s$^{-1}$
kpc$^{-1}$ for O and B stars and 10-20 km s$^{-1}$ kpc$^{-1}$ for
Cepheids). These ranges of values have allowed us to check how compatible
are the results independently obtained for O and B stars and Cepheids.

\item If we choose an incorrect set of free parameters ($m$, $i$,
$\varpi_\odot$, $\Theta(\varpi_\odot)$), results change, but not to a
large extent. This change is similar to the dispersion obtained when
solving the condition equations from the different simulated samples (see
Appendix \ref{ApenB} for details). Therefore, it will be difficult to
decide between 2- and 4-armed models.

\end{itemize}

From these results, in Appendix \ref{ApenB} we conclude that, with the
present available observational data, we are able to determine the
kinematic parameters of the galactic model proposed in this paper, though
we are not able to decide between a 2- or 4-armed Galaxy.

%
\section{Results}\label{results}

In this section we present the results obtained from our working samples.
To avoid those stars belonging to the Gould Belt, which can produce
important deviations in our results (see Paper I), we always considered
stars further than 0.6 kpc from the Sun. Following the results obtained in
Appendix \ref{ApenB}, the working distance intervals are 0.6 $< R <$ 2 kpc
for O and B stars and 0.6 $< R <$ 4 kpc for Cepheids.

A first set of results is presented in Table \ref{tab.res}, considering a
classical view of our galaxy (Lin et al. \cite{Lin et al.2}; Kerr \&
Lynden-Bell \cite{Kerr et al.}): we supposed a differential galactic
rotation ($a_\mathrm{r}$, $b_\mathrm{r}$) and a system with 2 spiral arms
and a pitch angle of $-6\degr$, the Sun placed at a galactocentric
distance of 8.5 kpc a the circular velocity at the Sun's position of 220
km s$^{-1}$.

\begin{table}
\caption[]{Resolution for the samples of O and B stars and Cepheids
           (SCS: short cosmic scale; LCS: large cosmic scale).
           Units: $U_\odot$, $V_\odot$, $W_\odot$, $\Pi_\mathrm{b}$,
           $\Theta_\mathrm{b}$ and $\sigma$ in km \mbox{s$^{-1}$};
           $a_\mathrm{r}$ in km \mbox{s$^{-1}$} \mbox{kpc$^{-1}$};
           $b_\mathrm{r}$ in km \mbox{s$^{-1}$} \mbox{kpc$^{-2}$};
           $\psi_\odot$ in degrees. $\chi^2/N$ is the  value of $\chi^2$
           divided by the number of equations minus the degrees of
           freedom. A cosmic dispersion of $(\sigma_U,\sigma_V,\sigma_W) =
           (8,8,5)$ km s$^{-1}$ and $(\sigma_U,\sigma_V,\sigma_W) =
           (13,13,6)$ km s$^{-1}$ was used for O and B stars and Cepheids, 
           respectively.}
\label{tab.res}
\begin{tabular}{crrr}
        \hline
        & O and B stars & \multicolumn{2}{c}{Cepheid stars} \\
        \hline
        & 0.6 $< R <$ 2 kpc & \multicolumn{2}{c}{0.6 $< R <$ 4 kpc} \\
        \hline
        & & SCS distances & LCS distances \\
        \hline
        \hline
$U_\odot$           &    8.8  $\pm$  0.7  &    6.5  $\pm$  1.2  &    8.3  $\pm$  1.2  \\
$V_\odot$           &   12.4  $\pm$  1.0  &   10.4  $\pm$  1.9  &    9.3  $\pm$  2.1  \\
$W_\odot$           &    8.4  $\pm$  0.5  &    5.7  $\pm$  0.7  &    7.0  $\pm$  0.8  \\
$a_\mathrm{r}$      & $-$1.3  $\pm$  1.0  & $-$8.1  $\pm$  1.2  & $-$4.5  $\pm$  1.1  \\
$b_\mathrm{r}$      & $-$0.8  $\pm$  1.5  &    1.8  $\pm$  0.8  & $-$0.1  $\pm$  0.8  \\
$\psi_\odot$        &   45.   $\pm$ 52.   &  282.   $\pm$ 20.   &  306.   $\pm$ 24.   \\
$\Pi_\mathrm{b}$    &    1.6  $\pm$  1.1  &    0.1  $\pm$  1.6  &    0.7  $\pm$  1.4  \\
$\Theta_\mathrm{b}$ &    2.1  $\pm$  1.5  &    4.9  $\pm$  1.7  &    4.8  $\pm$  1.7  \\
$f_\odot$           &    0.42 $\pm$  0.10 &    0.96 $\pm$  0.01 &    0.96 $\pm$  0.01 \\
        \hline
$\sigma$            &   12.11             &   11.57             &   11.77             \\
$\chi^2/N$          &    2.07             &    0.84             &    0.84    \\
        \hline
\end{tabular}  
\end{table}

Although we do not only present solutions from radial velocity or proper
motion data, we would like to remark that the parameters obtained when
solving these cases (and comparing with the combined solutions presented
in Table \ref{tab.res}) are compatible between themselves within the
errors bars. This is true for the sample of O and B stars and also in the
case of the sample of Cepheids (for both short and large cosmic scales).
Nevertheless, as we determined in Paper I, for this kind of kinematic
analysis it is advisable to solve a combined resolution, in order to
minimize the influence of the correlations present between the different
parameters. This is especially important in the present case, since some
correlations reach a large value (up to 0.8, though in the majority of
cases they do not exceed 0.3). In Appendix \ref{ApenB} we check that these
correlations do not impede our obtaining reliable results from our stellar
samples.

If we compare the results obtained from O and B stars and Cepheids, we
observe several discrepancies: differences up to 3.5 km s$^{-1}$
kpc$^{-1}$ in the $A$ Oort constant (derived from $a_\mathrm{r}$) and up
to 125$\degr$ for the phase of the spiral structure at the Sun's position.
We must keep in mind that, whereas this parameter (like
$\Omega_{\mathrm{p}}$ and $\varpi_{\mathrm{cor}}$) is independent of the
sample used, $\Pi_{\mathrm{b}}$, $\Theta_{\mathrm{b}}$ and $f_\odot$
depend on the cosmic dispersion of the sample (thus, they do not have the
same value for O and B stars and Cepheids).

As in Paper I, we found values of $\chi^2/N \approx 2$ for O and B stars.
We think that the difference from the expected value ($\chi^2/N \approx
1$) could be due to an underestimation of the errors in the photometric
distances and/or radial velocities for far stars. In the case of Cepheids
we found values which agree with the expected $\chi^2/N \approx 1$.

\begin{figure} 
  \resizebox{\hsize}{!}{\includegraphics{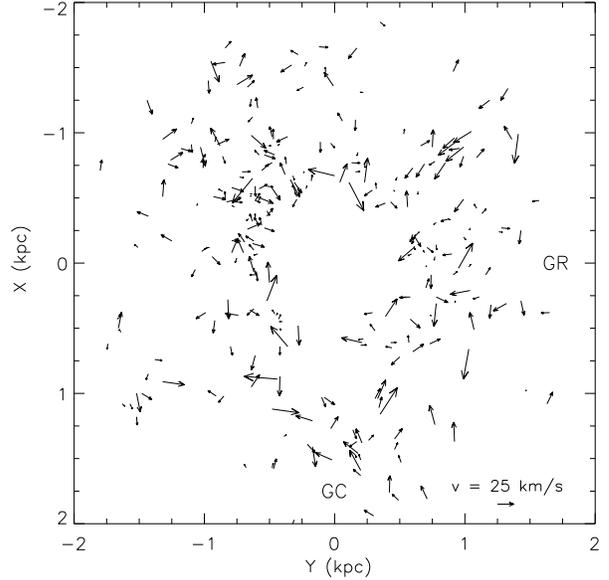}}
  \caption{Residual heliocentric space velocity vectors projected on the 
           galactic plane for O and B stars. GR stands for galactic
           rotation direction and GC for galactic center direction.}
  \label{fig.residus}
\end{figure}

When studying the residuals of the equations we realized that radial
velocity equations have a non-null average residual of about $-$(3-4) km
s$^{-1}$ for both O and B stars and Cepheids. In the case of O and B stars
(see Fig. \ref{fig.residus}) we can observe the peculiar motion of some OB
associations. The kinematics of those associations located near the
Sagittarius arm (e.g. \object{Sgr OB1}, \object{Ser OB1} and \object{Sct
OB2}, placed at $0\degr \la l \la 30\degr$) were studied by Mel'nik et al.
(\cite{Mel'nik et al.2}), who found that their motions are in general
agreement with what would be expected according to Lin's theory if these
stars are located within the corotation radius. But we found these stars
have a large residual, even taking into account in our model the galactic
spiral structure, possibly due to the fact that they are still reflecting
a motion peculiar to their birthplaces owing to their youth. Another
region with a high residual is that located in the direction of the open
clusters \object{$h$ and $\chi$ Per} (\object{NGC 869 and 884}), at $l
\approx 135\degr$, where we observe a group of stars (distributed in an
area of $10\degr$ in the sky) with a high residual directed towards the
Sun's position. Thus, our results indicate that these stars do not fit the
systematic velocity field defined by the whole sample. In this region of
the sky the associations \object{Per OB1} and \object{Cas OB6} are to be
found. The former is generally considered to include $h$ and $\chi$ Per
and surrounding O and B stars. However, the stars in Fig.
\ref{fig.residus} have distances $1.1 < R < 1.4$ kpc, whereas the open
clusters are located further on, at $R \approx$ 2-2.5 kpc (Schild
\cite{Schild}). In our initial sample (see Sect. \ref{samples}), we have
in this region a group of stars with distances $2.0 < R < 2.6$ kpc, most
of them belonging to \object{Per OB1} according to the membership list
done by Garmany \& Stencel (\cite{Garmany et al.}). But we did not use
these stars in our kinematic analysis, due to their large distance errors.
Garmany \& Stencel (\cite{Garmany et al.}) questioned if $h$ and $\chi$
Per really belong to \object{Per OB1} due to the fact that B main sequence
stars do not define a ZAMS at the distance of the open clusters. These
authors derived a distance modulus of 11.8 ($R = 2.3$ kpc) for \object{Per
OB1} and 11.9 ($R = 2.4$ kpc) for \object{Cas OB6}. More recently, Mel'nik
\& Efremov (\cite{Mel'nik et al.1}) studied the spatial distribution of O
and B stars within 3 kpc from the Sun and derived a new partition into OB
associations using Battinelli's (\cite{Battinelli}) modification of the
cluster analysis method. They found that \object{Per OB1} split into four
groups, two at $R \approx 1.6$ kpc and another two at $R \approx 1.9$ kpc.
Two of them are reliable at a 90\% confidence level (one at 1.6 kpc and
another at 1.9 kpc), with a mean heliocentric radial velocity of $-21.6$
km s$^{-1}$ and $-41.4$ km s$^{-1}$, respectively. The first association
found by Mel'nik \& Efremov might correspond to the group of stars we have
at $1.1 < R < 1.4$ kpc. In view of all that, the physical explanation of
the large negative radial residuals present in this region, even
considering the contribution from the spiral arm kinematics, is still
unresolved.

If we exclude those stars belonging to both regions from our calculations,
the average residual in radial velocity decreases to $-$(1-2) km s$^{-1}$.
For Cepheids, as mentioned above, the residuals have the same trend and
reach up to $-$(3-4) km s$^{-1}$, though in this case we cannot clearly
identify groups of stars with a common motion. Therefore, the mean
negative residual motion observed does not seem to be explained as the
peculiar motion of some groups of stars. Moreover, we found that, for both
samples, this residual motion apparently does not depend on the galatic
longitude of the stars.

Taking into account that Cepheids, with a large mean distance, show larger
residuals than O and B stars, we can suppose that this systematic trend
could be explained through the addition of a $K$-term in Eqs.
(\ref{vtot}), as developed in Paper I. Then, the following terms can be
added to Eqs. (\ref{eq.fit}):
\begin{eqnarray}
f_{11}^\mathrm{r} & = & R \cos^2 b \nonumber \\
f_{11}^\mathrm{l} & = & 0 \nonumber \\
f_{11}^\mathrm{b} & = & - R \sin b \cos b
\end{eqnarray}
\noindent which allow us to derive the parameter $a_{11} = K$. Table
\ref{tab.results} shows the results obtained in this way, and considering
different sets of free parameters (as proposed in Sect.
\ref{subsect.parameters}). The correlation coefficient between $K$ and the
other parameters is always low, being smaller than 0.4 for O and B stars
and 0.2 for Cepheids. As stated in Appendix \ref{ApenB}, a change in the
value of the free parameters ($m$, $i$, $\varpi_\odot$ and
$\Theta(\varpi_\odot)$) does not significantly alter the kinematic
parameters derived. In Table \ref{tab.results} we also show the results
when rejecting those stars belonging to the problematic regions mentioned
above in the case of O and B stars. We observe that there is not a great
difference in the results between these cases, which are always compatible
within the errors. So, we conclude that the observed peculiar motion of
some OB associations present in our sample does not alter the main
parameters of our model. In view of that, in the next section we discuss
the results obtained when considering all the stars, without rejecting
associations.

\begin{table}
\caption[]{Resolution by least squares fit considering different 
           values of the imposed galactic parameters $m, i, 
           \varpi_\odot$ and $\Theta(\varpi_\odot)$.
           Case A: $m = 2 , i = -6\degr, \varpi_\odot = 8.5$ kpc, 
           $\Theta(\varpi_\odot) =$ 220 km s$^{-1}$;
           Case D: $m = 4, i = -14\degr, \varpi_\odot = 7.1$ kpc, 
           $\Theta(\varpi_\odot) =$ 184 km s$^{-1}$.
           Units: $U_\odot$, $V_\odot$, $W_\odot$, $\Pi_\mathrm{b}$,
           $\Theta_\mathrm{b}$ and $\sigma$ in km \mbox{s$^{-1}$}; 
           $a_\mathrm{r}$, $K$, $A$, $B$ and $\Omega_{\mathrm{p}}$ in
           km \mbox{s$^{-1}$} \mbox{kpc$^{-1}$}; $b_\mathrm{r}$ in km
           \mbox{s$^{-1}$} \mbox{kpc$^{-2}$}; $\psi_\odot$ in degrees; 
           $\Delta \varpi_{\mathrm{cor}}= \varpi_\odot - 
           \varpi_{\mathrm{cor}}$ in kpc. $\chi^2/N$ is the value of
           $\chi^2$ divided by the number of equations minus the degrees
           of freedom. G1 and G2 stand for those groups of stars located
           in $0 < l < 50\degr$, $1 < R < 2$ kpc (29 stars) and $130 < l <
           140\degr$, $1 < R < 2$ kpc (14 stars), respectively.}
\label{tab.results}
\begin{tabular}{crrrrrrrr}
        \hline
	& \multicolumn{4}{c}{O and B stars with 0.6 $< R <$ 2 kpc} \\
        \hline
        \hline
        &\hspace{-3mm}Case A 
        &\hspace{-3mm}Case D 
        &\hspace{-3mm}Case D 
        &\hspace{-3mm}Case D \\
        \hline
        &\hspace{-3mm}
        &\hspace{-3mm}        
        &\hspace{-3mm}excluding G1 
        &\hspace{-3mm}excluding G2 \\
        \hline
        \hline
\hspace{-3mm}$U_\odot$           &\hspace{-3mm}9.2  $\pm$  0.7  
                                 &\hspace{-3mm}10.0  $\pm$  0.7  
                                 &\hspace{-3mm}9.2  $\pm$  0.8  
                                 &\hspace{-3mm}9.6  $\pm$  0.8  \\
\hspace{-3mm}$V_\odot$           &\hspace{-3mm}12.7  $\pm$  1.1  
                                 &\hspace{-3mm}12.7  $\pm$  1.0  
                                 &\hspace{-3mm}13.2  $\pm$  1.0  
                                 &\hspace{-3mm}12.3  $\pm$  1.1  \\
\hspace{-3mm}$W_\odot$           &\hspace{-3mm}8.3  $\pm$  0.5  
                                 &\hspace{-3mm}8.3  $\pm$  0.5  
                                 &\hspace{-3mm}8.3  $\pm$  0.5  
                                 &\hspace{-3mm}8.4  $\pm$  0.5  \\
\hspace{-3mm}$a_\mathrm{r}$      &\hspace{-3mm}$-$1.7  $\pm$  1.0  
                                 &\hspace{-3mm}$-$1.5  $\pm$  1.0  
                                 &\hspace{-3mm}$-$2.3  $\pm$  1.2  
                                 &\hspace{-3mm}$-$0.9  $\pm$  1.0  \\
\hspace{-3mm}$b_\mathrm{r}$      &\hspace{-3mm}$-$0.4  $\pm$  1.5  
                                 &\hspace{-3mm}0.5  $\pm$  1.3  
                                 &\hspace{-3mm}2.3  $\pm$  1.5  
                                 &\hspace{-3mm}0.0  $\pm$  1.4  \\
\hspace{-3mm}$K$                 &\hspace{-3mm}$-$3.2  $\pm$  0.7  
                                 &\hspace{-3mm}$-$2.8  $\pm$  0.7  
                                 &\hspace{-3mm}$-$1.4  $\pm$  0.7  
                                 &\hspace{-3mm}$-$2.4  $\pm$  0.7  \\
\hspace{-3mm}$\psi_\odot$        &\hspace{-3mm}20.   $\pm$ 53.   
                                 &\hspace{-3mm}329.   $\pm$ 47.   
                                 &\hspace{-3mm}315.   $\pm$ 32.   
                                 &\hspace{-3mm}8.   $\pm$ 52.   \\
\hspace{-3mm}$\Pi_\mathrm{b}$    &\hspace{-3mm}3.1  $\pm$  1.1  
                                 &\hspace{-3mm}2.6  $\pm$  1.0  
                                 &\hspace{-3mm}3.2  $\pm$  1.1  
                                 &\hspace{-3mm}2.8  $\pm$  1.1  \\
\hspace{-3mm}$\Theta_\mathrm{b}$ &\hspace{-3mm}2.0  $\pm$  1.3  
                                 &\hspace{-3mm}2.1  $\pm$  1.3  
                                 &\hspace{-3mm}3.2  $\pm$  1.5  
                                 &\hspace{-3mm}1.8  $\pm$  1.3  \\
\hspace{-3mm}$f_\odot$           &\hspace{-3mm}0.23 $\pm$  0.57 
                                 &\hspace{-3mm}0.35 $\pm$  0.25 
                                 &\hspace{-3mm}0.40 $\pm$  0.10 
                                 &\hspace{-3mm}0.19 $\pm$  0.77 \\
        \hline
\hspace{-3mm}$A$                 &\hspace{-3mm}13.8  $\pm$  0.5  
                                 &\hspace{-3mm}13.7  $\pm$  0.5  
                                 &\hspace{-3mm}14.1  $\pm$  0.6  
                                 &\hspace{-3mm}13.4  $\pm$  0.5  \\
\hspace{-3mm}$B$                 &\hspace{-3mm}$-$12.7  $\pm$  0.8  
                                 &\hspace{-3mm}$-$13.2  $\pm$  0.7  
                                 &\hspace{-3mm}$-$14.1  $\pm$  0.8  
                                 &\hspace{-3mm}$-$13.0  $\pm$  0.7  \\
\hspace{-3mm}$\Omega_{\mathrm{p}}$ 
                                 &\hspace{-3mm}45.   $\pm$ 14.   
                                 &\hspace{-3mm}33.   $\pm$  6.   
                                 &\hspace{-3mm}32.   $\pm$  3.   
                                 &\hspace{-3mm}36.   $\pm$  8.   \\
\hspace{-3mm}$\Delta \varpi_{\mathrm{cor}}$
                                 &\hspace{-3mm}3.6  $\pm$  1.6  
                                 &\hspace{-3mm}1.5  $\pm$  0.9  
                                 &\hspace{-3mm}1.1  $\pm$  0.6  
                                 &\hspace{-3mm}2.0  $\pm$  1.2  \\
        \hline
\hspace{-3mm}$\sigma$            &\hspace{-3mm}11.85             
                                 &\hspace{-3mm}11.88             
                                 &\hspace{-3mm}11.74             
                                 &\hspace{-3mm}11.85             \\
\hspace{-3mm}$\chi^2/N$          &\hspace{-3mm}1.89             
                                 &\hspace{-3mm}1.90
                                 &\hspace{-3mm}1.90
                                 &\hspace{-3mm}1.89              \\
        \hline
        \hline
	& \multicolumn{4}{c}{Cepheid stars with 0.6 $< R <$ 4 kpc} \\
        \hline
	& \multicolumn{2}{c}{\hspace{6mm}SCS distances} 
        & \multicolumn{2}{c}{\hspace{6mm}LCS distances} \\
        \hline
        \hline
        &\hspace{-3mm}Case A 
        &\hspace{-3mm}Case D 
        &\hspace{-3mm}Case A 
        &\hspace{-3mm}Case D \\
        \hline
        \hline
\hspace{-3mm}$U_\odot$           &\hspace{-3mm}6.5  $\pm$  1.2   
                                 &\hspace{-3mm}6.5  $\pm$  1.2  
                                 &\hspace{-3mm}8.4  $\pm$  1.2  
                                 &\hspace{-3mm}8.2  $\pm$  1.2  \\
\hspace{-3mm}$V_\odot$           &\hspace{-3mm}10.5  $\pm$  1.9  
                                 &\hspace{-3mm}11.0  $\pm$  2.1  
                                 &\hspace{-3mm}9.1  $\pm$  2.1  
                                 &\hspace{-3mm}10.1  $\pm$  2.3  \\
\hspace{-3mm}$W_\odot$           &\hspace{-3mm}5.7  $\pm$  0.7  
                                 &\hspace{-3mm}5.7  $\pm$  0.8  
                                 &\hspace{-3mm}7.0  $\pm$  0.8  
                                 &\hspace{-3mm}7.0  $\pm$  0.8  \\
\hspace{-3mm}$a_\mathrm{r}$      &\hspace{-3mm}$-$7.9  $\pm$  1.2  
                                 &\hspace{-3mm}$-$7.4  $\pm$  1.2  
                                 &\hspace{-3mm}$-$4.4  $\pm$  1.1  
                                 &\hspace{-3mm}$-$3.9  $\pm$  1.1  \\
\hspace{-3mm}$b_\mathrm{r}$      &\hspace{-3mm}1.6  $\pm$  0.8  
                                 &\hspace{-3mm}0.5  $\pm$  0.8  
                                 &\hspace{-3mm}0.1  $\pm$  0.8  
                                 &\hspace{-3mm}$-$0.9  $\pm$  0.8  \\
\hspace{-3mm}$K$                 &\hspace{-3mm}$-$0.8  $\pm$  0.5  
                                 &\hspace{-3mm}$-$1.0  $\pm$  0.5  
                                 &\hspace{-3mm}$-$1.2  $\pm$  0.5  
                                 &\hspace{-3mm}$-$1.2  $\pm$  0.5  \\
\hspace{-3mm}$\psi_\odot$        &\hspace{-3mm}286.   $\pm$ 21.   
                                 &\hspace{-3mm}284.   $\pm$ 39.   
                                 &\hspace{-3mm}310.   $\pm$ 21.   
                                 &\hspace{-3mm}321.   $\pm$ 43.   \\
\hspace{-3mm}$\Pi_\mathrm{b}$    &\hspace{-3mm}0.0  $\pm$  1.6  
                                 &\hspace{-3mm}$-$1.4  $\pm$  1.6  
                                 &\hspace{-3mm}0.9  $\pm$  1.3  
                                 &\hspace{-3mm}0.4  $\pm$  1.2  \\
\hspace{-3mm}$\Theta_\mathrm{b}$ &\hspace{-3mm}4.8  $\pm$  1.7  
                                 &\hspace{-3mm}2.4  $\pm$  1.7  
                                 &\hspace{-3mm}5.6  $\pm$  1.7  
                                 &\hspace{-3mm}2.7  $\pm$  1.7  \\
\hspace{-3mm}$f_\odot$           &\hspace{-3mm}0.96 $\pm$  0.01 
                                 &\hspace{-3mm}0.95 $\pm$  0.01 
                                 &\hspace{-3mm}0.96 $\pm$  0.01 
                                 &\hspace{-3mm}0.96 $\pm$  0.01 \\

        \hline
\hspace{-3mm}$A$                 &\hspace{-3mm}16.9  $\pm$  0.6  
                                 &\hspace{-3mm}16.6  $\pm$  0.6  
                                 &\hspace{-3mm}15.1  $\pm$  0.6  
                                 &\hspace{-3mm}14.9  $\pm$  0.6  \\
\hspace{-3mm}$B$                 &\hspace{-3mm}$-$13.7  $\pm$  0.4  
                                 &\hspace{-3mm}$-$13.2  $\pm$  0.4  
                                 &\hspace{-3mm}$-$13.0  $\pm$  0.4  
                                 &\hspace{-3mm}$-$12.5  $\pm$  0.4  \\
\hspace{-3mm}$\Omega_{\mathrm{p}}$ 
                                 &\hspace{-3mm}26.   $\pm$  3.   
                                 &\hspace{-3mm}23.   $\pm$  4.   
                                 &\hspace{-3mm}28.   $\pm$  3.   
                                 &\hspace{-3mm}27.   $\pm$  2.   \\
\hspace{-3mm}$\Delta \varpi_{\mathrm{cor}}$
                                 &\hspace{-3mm}0.0  $\pm$  1.0  
                                 &\hspace{-3mm}$-$0.7  $\pm$  1.2  
                                 &\hspace{-3mm}0.5  $\pm$  0.8  
                                 &\hspace{-3mm}0.2  $\pm$  0.6  \\
        \hline
\hspace{-3mm}$\sigma$            &\hspace{-3mm}11.56             
                                 &\hspace{-3mm}11.66             
                                 &\hspace{-3mm}11.70             
                                 &\hspace{-3mm}11.90             \\
\hspace{-3mm}$\chi^2/N$          &\hspace{-3mm}0.84
                                 &\hspace{-3mm}0.85
                                 &\hspace{-3mm}0.83
                                 &\hspace{-3mm}0.86              \\
        \hline
\end{tabular}
\end{table}

%
\section{Discussion}\label{discussion}

\subsection{Galactic rotation curve}

A difference in the $A$ Oort constant of approximately 3 km s$^{-1}$ 
kpc$^{-1}$ between solutions for O and B stars and Cepheids with short
cosmic scale distances is obtained, whereas large cosmic scale provides an
intermediate value:
\begin{eqnarray}
  A^\mathrm{OB} \approx 13.7\mbox{-}13.8 
  {\rm \; km \; s}^{-1} {\rm \; kpc}^{-1}
\nonumber \\
  A^\mathrm{Cep}_{\mathrm{Short}} \approx 16.6\mbox{-}16.9
  {\rm \; km \; s}^{-1} {\rm \; kpc}^{-1}
\nonumber \\
  A^\mathrm{Cep}_{\mathrm{Large}} \approx 14.9\mbox{-}15.1
  {\rm \; km \; s}^{-1} {\rm \; kpc}^{-1}
\end{eqnarray}

\noindent As can be seen in Appendix \ref{ApenB}, these differences cannot
be explained by the observational constraints present in both samples. A
similar discrepancy was found by Frink et al. (\cite{Frink et al.}), who
derived $A = 14.0 \pm 1.2$ km s$^{-1}$ kpc$^{-1}$ from a sample of O and B
stars, and $A = 15.8 \pm 1.6$ km s$^{-1}$ kpc$^{-1}$ from a sample of
Cepheids (in both cases the authors only considered those stars with a
heliocentric distance of less than 1 kpc). Nonetheless, our values for
Cepheids do not reach the higher values obtained by Glushkova et al.
(\cite{Glushkova et al.}; $A = 19.5 \pm 0.5$ km s$^{-1}$ kpc$^{-1}$),
Mishurov et al. (\cite{Mishurov et al.1}; $A = 20.9 \pm 1.2$ km s$^{-1}$
kpc$^{-1}$), Mishurov \& Zenina (\cite{Mishurov et al.2}; $A = 18.8 \pm
1.3$ km s$^{-1}$ kpc$^{-1}$) and L\'epine et al. (\cite{Lepine et al.}; $A
= 17.5 \pm 0.8$ km s$^{-1}$ kpc$^{-1}$). Pont et al. (\cite{Pont et al.1})
and Metzger et al. (\cite{Metzger et al.}), from radial velocities of
Cepheid stars, found values of $A = 15.9 \pm 0.3$ km \mbox{s$^{-1}$}
\mbox{kpc$^{-2}$} and $A = 15.5 \pm 0.4$ km \mbox{s$^{-1}$}
\mbox{kpc$^{-2}$}, respectively. More recently, and using proper motions
and distance calibration from Hipparcos data on Cepheid stars, Feast \&
Whitelock (\cite{Feast et al.1}) found a value of $A = 14.8 \pm 0.8$ km
\mbox{s$^{-1}$} \mbox{kpc$^{-1}$}. Using a similar sample, Feast et al.
(\cite{Feast et al.2}) found $A = 15.1 \pm 0.3$ km \mbox{s$^{-1}$}
\mbox{kpc$^{-1}$} from radial velocities. As we have mentioned in Sect.
\ref{results}, we found good coherence for radial velocity, proper motion
and combined solutions for both cosmic distance scales.

An attempt to explain these discrepancies was made by Olling \& Merrifield
(\cite{Olling et al.}), who studied the variation of the $A$ and $B$ Oort
{\it functions} and found that they significantly differ from the general
$\sim\Theta(\varpi)/\varpi$ dependence expected for a nearly flat rotation
curve. Inside the solar circle, the value of $A$ rises to 18 km s$^{-1}$
kpc$^{-1}$ for $\Delta\varpi = \varpi - \varpi_\odot \approx -0.5$ kpc,
disminishes to 16 km s$^{-1}$ kpc$^{-1}$ for $\Delta\varpi \approx -1.2$
kpc, and rises continuously for $\Delta\varpi \la -1.5$ kpc, to 19 km
s$^{-1}$ kpc$^{-1}$ for $\Delta\varpi \approx -2$ kpc (see Fig. 3 in
Olling \& Merrifield \cite{Olling et al.}). Contrary to that, beyond the
solar circle $A$ decreases to 10-12 km s$^{-1}$ kpc$^{-1}$, maintaining
this value in the interval $0 \la \Delta\varpi \la 2.5$ kpc. Although $A$
is a {\it local} parameter, describing the local shape of the rotation
curve, Olling \& Merrifield already pointed out that the discrepancies in
the results published in the literature may be produced by their
dependence on the galactocentric distance. Our O and B stars are
distributed along all the galactic longitudes, whereas the Cepheids are
predominantly concentrated inside the solar circle, with a peak in the
spatial distribution for $\Delta\varpi \approx -0.6$ kpc corresponding to
the Sagittarius-Carina arm (see Figs. \ref{fig.xy} and \ref{fig.xy.cef}).
For O and B stars we found a {\it classical} value of $A \approx 14$ km
s$^{-1}$ kpc$^{-1}$ (a value between 10-12 and 16-18 km s$^{-1}$
kpc$^{-1}$), whereas for Cepheids a value $A \approx$ 15-17 km s$^{-1}$
kpc$^{-1}$ (depending on the PL relation considered) was derived, in
agreement with Olling \& Merrifield's assumptions.

From the results obtained in Appendix \ref{ApenB}, we would expect
uncertainties in the second-order term of the rotation curve of $\approx
1.0$ km s$^{-1}$ kpc$^{-2}$ for O and B stars and $\approx 0.5$ km
s$^{-1}$ kpc$^{-2}$ in the case of Cepheids. Taking this and the results
in Table \ref{tab.results} into account, we can state that $b_\mathrm{r}$
does not differ from a null value more than 2 km s$^{-1}$ kpc$^{-2}$. Pont
et al. (\cite{Pont et al.1}) found a value $b_\mathrm{r} = -1.7 \pm 0.2$
km \mbox{s$^{-1}$} \mbox{kpc$^{-2}$}, whereas Feast et al. (\cite{Feast et
al.2}) found $b_\mathrm{r} = -1.6 \pm 0.2$ km \mbox{s$^{-1}$}
\mbox{kpc$^{-2}$}, both using radial velocities of Cepheid stars (the
latter with a Hipparcos distance calibration). A large positive value was
found by L\'epine et al. (\cite{Lepine et al.}), who derived $b_\mathrm{r}
= 5.0 \pm 1.0$ km \mbox{s$^{-1}$} \mbox{kpc$^{-2}$} from their sample of
Cepheid stars. As we will see in Sect. \ref{subsect.spiral}, these authors
also found a large value for the amplitude of the velocity component in
the galactic rotation direction due to the spiral potential
($\Theta_{\mathrm{b}}$). Without specific simulations, considering both
their observational data and resolution procedure, it is difficult to
guess how the correlations between both parameters can affect their
determination.

\subsection{Spiral structure}\label{subsect.spiral}

Appendix \ref{ApenB} demonstrates that the available observational data
allow the characterization of the galactic spiral structure, though the
biases and uncertainties on the parameters have to be taken into account
in the interpretation of the results. Furthermore, we realized that it is
very difficult to establish the correct number of spiral arms of the
Galaxy.

A first remarkable result is the fairly good coherence obtained for the
phase of the spiral structure at the Sun's position $\psi_\odot$ when
using different free parameters (cases A, D) or different samples compared
to the great discrepancies found in the literature (see below). We take
into consideration that $\Pi_{\mathrm{b}}$, $\Theta_{\mathrm{b}}$ and
$f_\odot$ depend on the cosmic dispersion and so they do not have the same
value for O and B stars and Cepheids.

\begin{figure}
  \resizebox{\hsize}{!}{\includegraphics{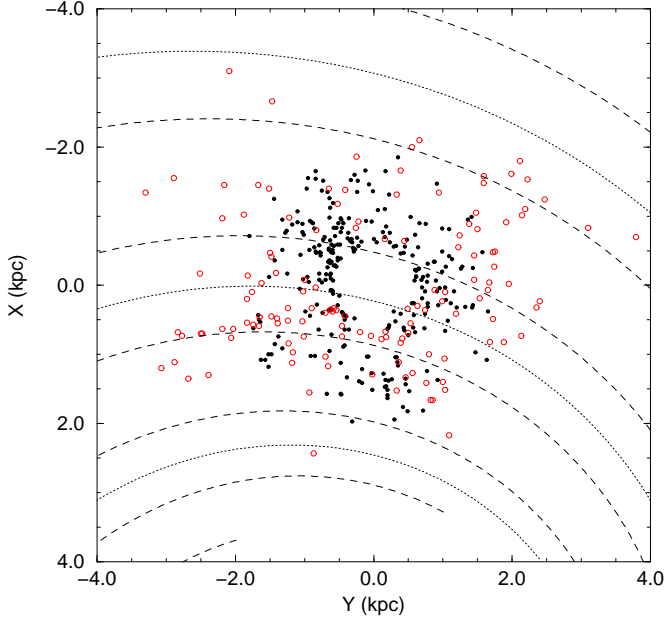}}
  \caption{Star distribution in the $X$-$Y$ galactic plane for O and B
           stars (filled circles) with 0.6 $< R <$ 2 kpc and Cepheids
           (empty circles) with 0.6 $< R <$ 4 kpc (short cosmic scale
           distances). Spiral arms were drawn considering $\psi_\odot = 
           330\degr$. Dotted lines show the center of the spiral arms
           ($\psi = 0\degr$) and dashed lines draw their approximate edges
           ($\psi = \pm90\degr$). Looking at this figure we must have in
           mind the possible drift between the position of the optical
           tracers in the spiral arms (i.e., our young stars) and the
           minimum of the spiral potential ($\psi = 0\degr$).}
  \label{fig.xy.OB+cef}
\end{figure}

From Fig. \ref{fig.xy.cef}, and assuming that Cepheids fairly trace the
center of the Sagittarius-Carina arm, we can infer a value of $\psi_\odot
\approx$ 250$\degr$ (the center of the inner visible spiral arm --the
Sagittarius-Carina arm-- at about 1 kpc from the Sun), depending on the
exact value of the interarm distance. Nevertheless, the density wave
theory predicts that the center of the visible arm (traced by its young
stars) does not coincide with the position of the spiral potential minimum
(Roberts \cite{Roberts1}, \cite{Roberts2}). We must bear in mind that the
kinematically derived value for $\psi_\odot$ informs us about the position
of the Sun with respect this potential minimum, not with respect the
visible arm. In Table \ref{tab.results} we found values inside the
interval:
\begin{equation}
  \psi_\odot \approx 284\mbox{-}20\degr
\end{equation}

\noindent We take into consideration that in the minimum of the spiral
potential (near the center of an arm) $\psi = 0\degr$, whereas in the
inner edge of the arm $\psi \approx 90\degr$ and in the outer edge $\psi
\approx -90\degr = 270\degr$. In the case of the phase of the spiral
structure at the Sun's position we expect a bias of $\Delta \psi_\odot =
\psi_\odot^{\mathrm{obtained}} - \psi_\odot^{\mathrm{real}} \approx
20\degr$ for O and B stars and $\Delta \psi_\odot \approx 0\degr$ for
Cepheids, with uncertainties of $\approx 25\degr$ and $\approx 75\degr$,
respectively (see Appendix \ref{ApenB}). These high uncertainties can
explain the range of values obtained. According to the value found for
$\psi_\odot$, the Sun is located between the center and the outer edge of
an arm, nearer to the former (see Fig. \ref{fig.xy.OB+cef}).

Our result stands in apparent contradiction to the classical mapping of
the spiral structure tracers (Schmidt-Kaler \cite{Schmidt-Kaler};
Elmegreen \cite{Elmegreen}), which locates the Sun in a middle position
between the inner arm (Sagittarius-Carina arm) and the outer one (Perseus
arm), that is, $\psi_\odot \approx 180\degr$. The local arm (Orion-Cygnus
arm) is normally considered as a local spur rather than a real arm. In our
model, as we supposed an interarm distance of approximately 3 kpc, the
local arm is also considered a local spur (see Sect.
\ref{subsect.parameters}). But the value we obtained for $\psi_\odot$
differs significantly from $180\degr$. If we consider a difference of
30-$100\degr$ between the optical tracers and the potential minimum of the
spiral arm, we notice that our $\psi_\odot$ value is in good agreement
with the picture of the spiral structure that emerges from the spatial
distribution of stars in Fig. \ref{fig.xy.OB+cef}, with the inner spiral
arm at approximately 1 kpc from the Sun. A similar result was obtained by
Mel'nik et al. (\cite{Mel'nik et al.2}), who found that nearly 70\% of the
stars in the OB associations of the Sagittarius-Carina arm have a residual
motion (after correcting their heliocentric velocities for solar motion
and galactic rotation) in the direction opposite to the galactic rotation,
as one would expect for those stars between the inner edge and the center
of the arm. Then, they also found a shift between the optical position of
the visible arm (traced by young stars) and the minimum of the spiral
potential. Therefore, from our results we conclude that the Sun is placed
relatively near the potential minimum of the Sagittarius-Carina arm, and
the Perseus arm is located far away, at about 2.5 kpc from the Sun.

Our range of values for $\psi_\odot$ includes that obtained by Cr\'ez\'e
\& Mennessier (\cite{Creze et al.}) from a sample of O-B3 stars,
$\psi_\odot = 352 \pm 30\degr$. Later, Mennessier \& Cr\'ez\'e
(\cite{Mennessier at al.})  found from a sample of O-B3 stars a value of
$\psi_\odot \approx 90\degr$, which locates the Sun near the inner edge of
an arm. Other authors obtained contradictory results. G\'omez \&
Mennessier (\cite{Gomez et al.}) found the Sun's position near the edge of
an arm from several samples of stars from FK4 and FK4 Supplement
Catalogues. Byl \& Ovenden (\cite{Byl et al.}) and Comer\'on \& Torra
(\cite{Comeron et al.1}) obtained values of $\psi_\odot = 165 \pm 1\degr$
and $\psi_\odot = 135 \pm 18\degr$ respectively, from samples of O and B
stars. Mishurov et al.  (\cite{Mishurov et al.1}) found $\psi_\odot = 290
\pm 16\degr$ from a sample of Cepheid stars with radial velocities.
Mishurov \& Zenina (\cite{Mishurov et al.2})  found $\psi_\odot = 322 \pm
9\degr$ (supposing $m = 2$ and $\varpi_\odot = 7.5$ kpc), from the same
sample of Cepheid stars, but also including the Hipparcos proper motions.
When these authors supposed $m = 4$ they found $\psi_\odot = 340 \pm
9\degr$. More recently, Rastorguev et al. (\cite{Rastorguev et al.}) found
$\psi_\odot = 274 \pm 22\degr$ from a sample of 55 open clusters younger
than 40 Myr and 67 Cepheids with periods smaller than 9 days, all of them
at $R < 4$ kpc. We can see that there is poor agreement between the
results published in the literature, though the last ones (the only ones
with Hipparcos data) are all included in our range of possible values for
$\psi_\odot$.

The velocity amplitudes due to the spiral perturbation obtained are less
than 4 km s$^{-1}$ for O and B stars ($\Pi_{\mathrm{b}} \approx 3$ km
s$^{-1}$, $\Theta_{\mathrm{b}} \approx$ 2-3 km s$^{-1}$) and 6 km s$^{-1}$
for Cepheid stars ($\Pi_{\mathrm{b}} \approx -1$-1 km s$^{-1}$,
$\Theta_{\mathrm{b}} \approx$ 2-6 km s$^{-1}$). The difference between
results for O and B stars and Cepheids is explained by Lin's theory as a
result of their slight dependence on the galactocentric distance and the
different cosmic dispersion for both samples. Mishurov et al.
(\cite{Mishurov et al.1}) found $\Pi_{\mathrm{b}} = 6.3 \pm 2.4$ km
s$^{-1}$ and $\Theta_{\mathrm{b}} = 4.4 \pm 2.4$ km s$^{-1}$ from their
radial velocity data of Cepheid stars within 4 kpc from the Sun, supposing
a 2-armed spiral pattern.  Mel'nik et al. (\cite{Mel'nik et al.3}) found
$\Pi_{\mathrm{b}} = 6.4 \pm 1.2$ km s$^{-1}$ and $\Theta_{\mathrm{b}} =
2.4 \pm 1.2$ km s$^{-1}$ from Cepheids within 3 kpc from the Sun.  
Mishurov \& Zenina (\cite{Mishurov et al.2}) found $\Pi_{\mathrm{b}} = 3.3
\pm 1.6$ km s$^{-1}$ and $\Theta_{\mathrm{b}} = 7.9 \pm 2.0$ km s$^{-1}$
for $m = 2$ and $\Pi_{\mathrm{b}} = 3.5 \pm 1.7$ km s$^{-1}$ and
$\Theta_{\mathrm{b}} = 7.5 \pm 1.8$ km s$^{-1}$ for $m = 4$. From the same
sample, L\'epine et al. (\cite{Lepine et al.}) found
$\Pi_{\mathrm{b}}^{m=2} = 0.4 \pm 3.0$ km s$^{-1}$,
$\Theta_{\mathrm{b}}^{m=2} = 14.0 \pm 3.0$ km s$^{-1}$ and
$\Pi_{\mathrm{b}}^{m=4} = 0.8 \pm 3.3$ km s$^{-1}$,
$\Theta_{\mathrm{b}}^{m=4} = 10.9 \pm 2.9$ km s$^{-1}$. Their 2+4-armed
model yields large values for $\Theta_{\mathrm{b}}$, implying a large
value for the ratio between the spiral potential and the axisymmetric
galactic field, much greater than that 5-10\% normally accepted. According
to the results obtained in Appendix \ref{ApenB}, the present observational
uncertainties, biases and the correlations involved in the resolution
procedure cannot completely explain the discrepancies among the values
found in the literature. They can be attributed to our poor knowledge of
the real wave harmonic structure of the galactic spiral pattern or to the
approximation performed in the linear density-wave theory.

The derived angular rotation velocity of the spiral pattern is:
\begin{equation}
  \Omega_{\mathrm{p}} \approx 30 {\rm \; km \; s}^{-1} {\rm \; kpc}^{-1}
\end{equation}

\noindent though there is a great dispersion around this value in our
results from different samples and cases (dispersion of 2-7 km s$^{-1}$
kpc$^{-1}$, but up to 15 km s$^{-1}$ kpc$^{-1}$ in one extreme case). This
is an expected dispersion, taking into account the results obtained in
Appendix \ref{ApenB} (standard deviation of 2-5 km s$^{-1}$ kpc$^{-1}$ for
O and B stars, but up to 15 km s$^{-1}$ kpc$^{-1}$ for $\psi_\odot \approx
270\degr$; for Cepheids dispersions are higher, of about 10-20 km s$^{-1}$
kpc$^{-1}$). Such a value places the Sun very near the corotation circle
and is not in agreement with the classical $\Omega_{\mathrm{p}} \approx
13.5$ km s$^{-1}$ kpc$^{-1}$ proposed by Lin et al. (\cite{Lin et al.2}).
Nevertheless, other studies show results similar to ours. Avedisova
(\cite{Avedisova}) obtained $\Omega_{\mathrm{p}} = 26.8 \pm 2$ km s$^{-1}$
kpc$^{-1}$ from the spatial distribution of young objects of different
ages in the Sagittarius-Carina arm. More recently, Amaral \& L\'epine
(\cite{Amaral et al.}), working with a selection of young members of the
open cluster catalogue by Mermilliod (\cite{Mermilliod}), obtained a value
of $\Omega_{\mathrm{p}} \approx$ 20-22 km s$^{-1}$ kpc$^{-1}$.  Mishurov
et al. (\cite{Mishurov et al.1}) and Mishurov \& Zenina (\cite{Mishurov et
al.2}) used their sample of Cepheid stars and found $\Omega_{\mathrm{p}} =
28.1 \pm 2.0$ km s$^{-1}$ kpc$^{-1}$ and $\Omega_{\mathrm{p}} \approx
27.7$ km s$^{-1}$ kpc$^{-1}$, respectively. L\'epine et al. (\cite{Lepine
et al.}) found $\Omega_{\mathrm{p}}^{m=2} \approx
\Omega_{\mathrm{p}}^{m=4} \approx 26.5$ km s$^{-1}$ kpc$^{-1}$ from their
sample of Cepheid stars. Rastorguev et al. (\cite{Rastorguev et al.})
thought that the evaluation of $\Omega_{\mathrm{p}}$ from kinematical data
alone cannot be resolved. But our simulations (see Appendix \ref{ApenB})
seem to indicate that, at least, the tendency to find high values of
$\Omega_{\mathrm{p}}$ is confirmed, though there is still an uncertainty
of about 5-10 km s$^{-1}$ kpc$^{-1}$ in its value. These large values for
$\Omega_{\mathrm{p}}$ can explain in a natural way the presence of a gap
in the galactic gaseous disk (see Kerr \cite{Kerr} and Burton
\cite{Burton2} for observational evidences and L\'epine et al.
\cite{Lepine et al.} for simulated results), since they placed the Sun
near the corotation circle, where the gas is pumped out under the
influence of the spiral potential.

\subsection{The $K$-term}

In relation to the $K$-term, we found good agreement between both O and B
stars and Cepheids. A value of $K \approx -$(1-3) km s$^{-1}$ kpc$^{-1}$
is found in all cases, confirming an apparent compression of the solar
neighbourhood of up to 3-4 kpc. In our opinion, what most clearly proves
the existence of a non-null value of $K$ is that it is independently found
from the samples of O and B stars and Cepheids, which have a very
different spatial distribution, an independent distance estimation and a
different way of deriving their radial velocities. As we can see comparing
Tables \ref{tab.res} and \ref{tab.results}, the inclusion of the $K$-term
does not substantially modify the other model parameters derived by least
squares fit.

It is very difficult to find in the literature other estimations of $K$ at
these large heliocentric distances, since in the majority of cases authors
consider an axisymmetric rotation curve. But some authors have pointed out
the persistence of a residual in the radial velocity equations. Comer\'on
\& Torra (\cite{Comeron et al.2}) found $K = -1.9 \pm 0.5$ km s$^{-1}$
kpc$^{-1}$ from O-B5.5 stars and $K = -1.3 \pm 0.9$ km s$^{-1}$ kpc$^{-1}$
from B6-A0 stars with $R < 1.5$ kpc. The radial velocity residual for
Cepheids was first recognised by Stibbs (\cite{Stibbs}).  Pont et al.
(\cite{Pont et al.1}) studied three possible origins for it: a statistical
effect, an intrinsic effect in the measured radial velocities for Cepheids
($\gamma$-velocities) and a real dynamical effect. They finally suggested
that it could be due to a non-axisymmetric motion produced by a central
bar of $\approx 5$ kpc in extent. Metzger et al. (\cite{Metzger et al.})
found a residual of $\approx-3$ km s$^{-1}$ for a sample of Cepheids when
considering an axisymmetric galactic rotation. They concluded that it
might be due to the influence of spiral structure, not included in their
model. However, in the present work we found a non-null $K$ value even
taking into account spiral arm kinematics. An understanding of the
physical explanation of the $K$-term requires further study.

%
\section{Conclusions}

Hipparcos astrometric data were used to derive the spiral structure of the
Galaxy in the solar neighbourhood. We considered two different samples of
stars as tracers of this structure. In the sample of O and B stars the
astrometric data were complemented with a careful compilation of radial
velocities and Str\"omgren photometry, providing reliable distances and
spatial velocities for these stars. In the sample of Cepheid stars we used
two period-luminosity relations, one considering a short cosmic scale
(Luri \cite{Luri}) and another one standing for a large cosmic scale
(Feast \& Catchpole \cite{Feast et al.0}).
	
A kinematic model of our galaxy that takes into account solar motion,
differential galactic rotation, spiral arm kinematics and a $K$-term was
adopted. The model parameters were derived via a classical weighted least
squares fit. The robustness of our results was checked by means of careful
insimulations.

A galactic rotation curve with a value of $A$ Oort constant of 13.7-13.8
km s$^{-1}$ kpc$^{-1}$ was found for the sample of O and B stars. In the
case of Cepheid stars, we found $A =$ 14.9-16.9 km s$^{-1}$ kpc$^{-1}$,
depending on the case and the cosmic scale chosen. We confirmed the
discrepancies appearing in $A$ when samples with a different distance
horizon were used (Olling \& Merrifield \cite{Olling et al.}). For both
cosmic scales we found an acceptable coherence between radial velocity,
proper motion and combined solutions. Concerning the second-order term of
the rotation curve, we always found a low value, compatible with zero.

The study of the residuals for radial velocity data made it evident that a
$K$-term was needed, which was found to be $K = -$(1.4-3.2) km s$^{-1}$
kpc$^{-1}$ for O and B stars and $K = -$(0.8-1.2) km s$^{-1}$ kpc$^{-1}$
for Cepheids. Although small, an apparent compression motion seems to
exist in the galactic local neighbourhood for heliocentric distances up to
4 kpc, though its physical mechanism is still unknown.

The phase of the spiral structure at the Sun's position obtained
($\psi_\odot = 284$-20$\degr$) places it between the center and the outer
edge of an arm. This is in good agreement with the spatial distribution of
Cepheid stars. The angular rotation velocity of the spiral structure was
found to be $\Omega_{\mathrm{p}} \approx$ 30 km s$^{-1}$ kpc$^{-1}$, which
places the Sun near the corotation circle.

%
\begin{acknowledgements}

This work has been supported by the CICYT under contracts ESP 97-1803 and
AYA 2000-0937. DF acknowledges the FRD grant from the Universitat de
Barcelona (Spain).

\end{acknowledgements}

%
\appendix

%
\section{systematic velocity components in the proposed galactic
model}\label{ApenA}

In this appendix we show the expressions of the velocity components in the
three systematic contributions considered in our galactic model: solar
motion, differential galactic rotation and spiral arm kinematics.

\subsection{Solar motion}

A star with galactic longitude $l$ and galactic latitude $b$ has the
following radial and tangential velocity components owing to solar proper
motion:
\begin{eqnarray}\label{eq.v1}
  v_{\mathrm{{r_1}}} & = & - U_\odot \cos l \cos b - V_\odot \sin l \cos b
  - W_\odot \sin b
  \nonumber \\
  v_{\mathrm{{l_1}}} & = & U_\odot \sin l - V_\odot \cos l
  \nonumber \\
  v_{\mathrm{{b_1}}} & = & U_\odot \cos l \sin b + V_\odot \sin l \sin b
  - W_\odot \cos b
\end{eqnarray}
where $U_\odot$, $V_\odot$ and $W_\odot$ are the components of the solar
motion in galactic coordinates.

\subsection{Galactic rotation}

We consider axisymmetric differential rotation of our galaxy, with a
rotation curve that can be developed in the solar neighbourhood as:
\begin{eqnarray}\label{vrot}
  \Theta(\varpi) & \approx & \Theta(\varpi_\odot) + 
  \left( \frac {\partial \Theta}{\partial \varpi} \right)_\odot 
  \Delta\varpi +
  \frac {1}{2} \left( \frac {\partial^2 \Theta}{\partial
  \varpi^2} \right)_\odot \Delta\varpi^2
  \nonumber \\
  & \equiv & \Theta(\varpi_\odot) + a_{\mathrm{r}} \Delta\varpi
  + b_{\mathrm{r}} \Delta\varpi^2
\end{eqnarray}
where $\Delta\varpi = \varpi - \varpi_\odot$ ($\varpi_\odot$ is the
galactocentric distance of the Sun) and $\Theta(\varpi_\odot)$ is the
circular velocity at the Sun's position. We note that $a_{\mathrm{r}}$
allows us to calculate the $A$ Oort constant in the Sun's vicinity: 
\begin{equation}\label{A}
  A = \frac {1}{2} \left[ \frac {\Theta(\varpi_\odot)}{\varpi_\odot} - 
  \left( \frac {\partial \Theta}{\partial \varpi} \right)_\odot \right]
  = \frac {1}{2} \left[ \frac {\Theta(\varpi_\odot)}{\varpi_\odot} -
  a_{\mathrm{r}} \right]
\end{equation}

The radial and tangential velocity components of a star in the galactic
plane due to differential galactic rotation are:
\begin{eqnarray}\label{eq.v2}
  v_{\mathrm{r_2}} & = & \Theta(\varpi_\odot) \left[\sin(l+\theta) -
  \sin l\right] \cos b \nonumber \\ 
  & & + a_{\mathrm{r}} \Delta\varpi \sin(l+\theta) \cos b
  \nonumber \\
  & & + b_{\mathrm{r}} \Delta\varpi^{\mathrm{2}} \sin(l+\theta) \cos b
  \nonumber \\
  \nonumber \\
  v_{\mathrm{l_2}} & = & \Theta(\varpi_\odot) \left[\cos(l+\theta) -
  \cos l\right] \nonumber \\ 
  & & + a_{\mathrm{r}} \Delta\varpi \cos(l+\theta)
  \nonumber \\
  & & + b_{\mathrm{r}} \Delta\varpi^{\mathrm{2}} \cos(l+\theta)
  \nonumber \\
  \nonumber \\
  v_{\mathrm{b_2}} & = & - \Theta(\varpi_\odot) \left[\sin(l+\theta) -
  \sin l\right] \sin b \nonumber \\ 
  & & - a_{\mathrm{r}} \Delta\varpi \sin(l+\theta) \sin b
  \nonumber \\
  & & - b_{\mathrm{r}} \Delta\varpi^{\mathrm{2}} \sin(l+\theta) \sin b
\end{eqnarray}
where $\theta$ is the galactocentric longitude of the star.

\subsection{Spiral structure kinematics}

Lin's theory (Lin \& Shu \cite{Lin et al.1}; Lin et al. \cite{Lin et  
al.2}; see also Rohlfs \cite{Rohlfs}) assumes a spiral potential of the 
form:
\begin{equation}\label{pot}
  V_{\mathrm{b}} = {\cal A} \cos \psi
\end{equation}

\noindent (${\cal A}$ is the amplitude -- ${\cal A} < 0$ -- and $\psi$ is
the phase of the density wave) which disturbs the axially symmetric
gravitational potential. The shape of the spiral arms is well represented
by a logarithmic spiral:
\begin{equation}\label{esp}
  q(\varpi,\theta,t) = q \, e^{i \psi(\varpi,\theta,t)}
\end{equation}

\noindent The amplitude $q$ is a slowly varying function of $\varpi$ and
the phase of the spiral structure can be related to the phase at the Sun's
position by the following expression:
\begin{eqnarray}\label{fase}
  \psi(\varpi,\theta,t) = \psi_\odot(t) + m (\Omega_{\mathrm{p}} t -
  \theta) + \frac {m \ln \frac {\varpi}{\varpi_\odot(t)}}{\tan i}
  \Rightarrow \nonumber \\
  \Rightarrow \psi(\varpi,\theta,t=0) = \psi_\odot + m \theta + \frac
  {m \ln \frac {\varpi}{\varpi_\odot}} {\tan i}
\end{eqnarray}

\noindent where $\Omega_{\mathrm{p}}$ is the angular rotation velocity of
the spiral pattern, $m$ the number of spiral arms and $i$ the pitch angle
(for trailing spiral arms, $i < $ 0). The phase of the spiral structure at
the Sun's position and the pitch angle can be determined from optical and
radio indicators. Nevertheless, we point out that the maximum in the
distribution of spiral arm tracers (position of the observed spiral arms)
may be shifted in relation to the minimum in the perturbation potential
(defined as $\psi = 0\degr$; see Roberts \cite{Roberts1}).

The mean peculiar velocities due to the spiral arm perturbations on the
velocity field are, in the gas approximation, the following:
\begin{eqnarray}\label{comp}
  \Pi_1 & = & \frac {k {\cal A}}{\kappa} \frac {\nu}{1 - \nu^2 + x} \cos\psi
  \equiv \Pi_{\mathrm{b}} \cos\psi \nonumber \\
  \Theta_1 & = & - \frac {1}{2} \frac{k {\cal A} \varpi}{\Theta} 
  \frac {1}{1 - \nu^2 + x} \sin\psi 
  \equiv -\Theta_{\mathrm{b}} \sin\psi
\end{eqnarray}

\noindent $\Pi_1$ is positive towards the galactic anti-center and
$\Theta_1$ is positive towards the galactic rotation. In tightly wound
spirals (i.e., those with $|\tan i| \ll 1$) the amplitudes
$\Pi_{\mathrm{b}}$ and $\Theta_{\mathrm{b}}$ are slowly varying quantities
with the galactocentric distance. $k$ is the radial wave number (for
trailing spiral arms, $k < $ 0):
\begin{equation}\label{k}
  k = \frac {d}{d \varpi} 
  \left( \frac {m \ln \frac {\varpi}{\varpi_\odot}}{\tan i} \right)
  = \frac {m}{\varpi \tan i}
\end{equation}

\noindent $\nu$ is the dimensionless rotation frequency of the spiral
structure, expressed in terms of the epicyclic frequency ($\kappa$):
\begin{equation}\label{nu}
  \nu = \frac {m}{\kappa} \left( \Omega_{\mathrm{p}} - 
  \frac {\Theta}{\varpi} \right)
\end{equation} 

\noindent (notice that $\nu < 0$ in the region with $\Omega_{\mathrm{p}} <
\Omega = \Theta/\varpi$, i.e. inner to the corotation circle). Furthermore:
\begin{equation}\label{kappa}
  \kappa^2 = \frac {2 \Theta^2}{\varpi^2} \left( 1 + 
  \frac {\varpi}{\Theta} \frac {d \Theta}{d \varpi} \right)
\end{equation} 

\noindent and $x$ is the stability Toomre's number (Toomre \cite{Toomre})
defined as:
\begin{equation}
  x = \frac {k^2 a^2_{\mathrm{o}}}{\kappa^2}
\end{equation}

\noindent where $a_{\mathrm{o}}$ is the velocity dispersion of the gas
particles. Since the velocity amplitudes $\Pi_{\mathrm{b}}$ and
$\Theta_{\mathrm{b}}$ depend on this velocity dispersion, we introduce a
dimensionless parameter ($f_\odot$) that relates the velocity amplitudes
of the Sun to those of the sample stars:
\begin{eqnarray}\label{f}
   \Pi_{\mathrm{b \odot}} = \frac {1 - \nu^2 + 
      x_{\mathrm{stars}}}{1 - \nu^2 + x_\odot} 
      \Pi_{\mathrm{b}} \equiv f_\odot \Pi_{\mathrm{b}} \nonumber \\
   \Theta_{\mathrm{b \odot}} = \frac {1 - \nu^2 + 
      x_{\mathrm{stars}}}{1 - \nu^2 + x_\odot}   
      \Theta_{\mathrm{b}} \equiv f_\odot \Theta_{\mathrm{b}}  
\end{eqnarray}

The so-called Lindblad resonances are defined as:
\begin{equation}\label{Linres}
  \Omega_{\mathrm{p}} = \Omega \pm \frac {\kappa}{m}
\end{equation}

\noindent where the $-$ sign corresponds to the inner resonance and the
$+$ sign to the outer one. In the region between both resonances ($|\nu| <
1$), $\Theta_{\mathrm{b}}$ is always positive and $\Pi_{\mathrm{b}}$ has a
sign that depends on the sign of $\nu$.

The amplitude of the spiral potential can be expressed as:
\begin{equation}
  {\cal A} = \frac {\kappa \Pi_{\mathrm{b}}}{k} \frac {1 - \nu^2 + x}{\nu}
\end{equation}

\noindent The maximum value of the radial force owing to this spiral
potential is:
\begin{equation}
  F_{{\mathrm{r}}1}^{\mathrm{max}} = k |{\cal A}|
\end{equation}

\noindent whereas the radial force due to the axisymmetric field is:
\begin{equation}
  F_{{\mathrm{r}}0} = \frac {dV_0}{d\varpi} \approx \Omega_\odot^2
  \varpi_\odot
\end{equation}

\noindent Therefore, the ratio between both quantities is:
\begin{equation}
  f_{\mathrm{r}} = \frac
  {F_{{\mathrm{r}}1}^{\mathrm{max}}}{F_{{\mathrm{r}}0}} \approx 
  \frac {\kappa \Pi_{\mathrm{b}}}{\Omega_\odot^2 \varpi_\odot}
  \frac {1 - \nu^2 + x}{\nu}
\end{equation}

The contributions in the radial and tangential velocity components of a
star due to the spiral arm perturbation velocities $\Pi_1$ and $\Theta_1$
are the following:
\begin{eqnarray}\label{eq.v3}
  v_{\mathrm{{r_3}}}
  & & = - \Pi_1 \cos (l + \theta) \cos b + \Pi_{1\odot} \cos l \cos b
  \nonumber \\
  & & + \Theta_1 \sin (l + \theta) \cos b - \Theta_{1\odot} \sin l \cos b
  \nonumber \\ 
  & & = - \Pi_{\mathrm{b}} \cos \psi_\odot \nonumber \\
  & & \times \left(
  \cos \left[ m \left( \theta - 
  \frac {\ln \frac {\varpi}{\varpi_\odot}}{\tan i} \right) \right]
  \cos(l + \theta) - f_\odot \cos l \right) \cos b \nonumber \\
  & & - \Pi_{\mathrm{b}} \sin \psi_\odot 
  \sin \left[ m \left(\theta -
  \frac {\ln \frac {\varpi}{\varpi_\odot}}{\tan i} \right) \right]
  \cos(l + \theta) \cos b \nonumber \\
  & & - \Theta_{\mathrm{b}} \sin \psi_\odot \nonumber \\
  & & \times \left(
  \cos \left[ m \left( \theta -
  \frac {\ln \frac {\varpi}{\varpi_\odot}}{\tan i} \right) \right]
  \sin(l + \theta) - f_\odot \sin l \right) \cos b \nonumber \\
  & & + \Theta_{\mathrm{b}} \cos \psi_\odot
  \sin \left[ m \left(\theta -
  \frac {\ln \frac {\varpi}{\varpi_\odot}}{\tan i} \right) \right]
  \sin(l + \theta) \cos b
  \nonumber \\
  \nonumber \\
  v_{\mathrm{{l_3}}}
  & & = \Pi_1 \sin (l + \theta) - \Pi_{1\odot} \sin l+
  \nonumber \\
  & & \Theta_1 \cos (l + \theta) - \Theta_{1\odot} \cos l
  \nonumber \\
  & & = \Pi_{\mathrm{b}} \cos \psi_\odot \nonumber \\
  & & \times \left(
  \cos \left[ m \left( \theta - 
  \frac {\ln \frac {\varpi}{\varpi_\odot}}{\tan i} \right) \right]
  \sin(l + \theta) - f_\odot \sin l \right) \nonumber \\ 
  & & + \Pi_{\mathrm{b}} \sin \psi_\odot 
  \sin \left[ m \left( \theta -
  \frac {\ln \frac {\varpi}{\varpi_\odot}}{\tan i} \right) \right]
  \sin(l + \theta) \nonumber \\
  & & - \Theta_{\mathrm{b}} \sin \psi_\odot \nonumber \\
  & & \times \left(
  \cos \left[ m \left( \theta -
  \frac {\ln \frac {\varpi}{\varpi_\odot}}{\tan i} \right) \right]
  \cos(l + \theta) - f_\odot \cos l \right) \nonumber \\
  & & + \Theta_{\mathrm{b}} \cos \psi_\odot 
  \sin \left[ m \left( \theta -
  \frac {\ln \frac {\varpi}{\varpi_\odot}}{\tan i} \right) \right]
  \cos(l + \theta)
  \nonumber \\
  \nonumber \\
  v_{\mathrm{{b_3}}}
  & & = \Pi_1 \cos (l + \theta) \sin b - \Pi_{1\odot} \cos l \sin b
  \nonumber \\
  & & - \Theta_1 \sin (l + \theta) \sin b + \Theta_{1\odot} \sin l \sin b
  \nonumber \\ 
  & & = \Pi_{\mathrm{b}} \cos \psi_\odot \nonumber \\
  & & \times \left(
  \cos \left[ m \left( \theta - 
  \frac {\ln \frac {\varpi}{\varpi_\odot}}{\tan i} \right) \right]
  \cos(l + \theta - f_\odot \cos l \right) \sin b \nonumber \\
  & & + \Pi_{\mathrm{b}} \sin \psi_\odot 
  \sin \left[ m \left(\theta -
  \frac {\ln \frac {\varpi}{\varpi_\odot}}{\tan i} \right) \right]
  \cos(l + \theta) \sin b \nonumber \\
  & & + \Theta_{\mathrm{b}} \sin \psi_\odot \nonumber \\
  & & \times \left(
  \cos \left[ m \left( \theta -
  \frac {\ln \frac {\varpi}{\varpi_\odot}}{\tan i} \right) \right]
  \sin(l + \theta) - f_\odot \sin l \right) \sin b \nonumber \\
  & & - \Theta_{\mathrm{b}} \cos \psi_\odot
  \sin \left[ m \left(\theta -
  \frac {\ln \frac {\varpi}{\varpi_\odot}}{\tan i} \right) \right]
  \sin(l + \theta) \sin b \nonumber \\
\end{eqnarray}

\subsection{Systematic velocity field in the proposed galactic model}
The systematic radial and tangential velocity components of a star in our
galactic model are given by:
\begin{eqnarray}
  v_{\mathrm{r}} = v_{\mathrm{r_1}} + v_{\mathrm{r_2}} + v_{\mathrm{r_3}}
  = \sum\limits_{j=1}^{10} a_j f_j^{\mathrm{r}}(R,l,b)
  \nonumber \\
  v_{\mathrm{l}} = v_{\mathrm{l_1}} + v_{\mathrm{l_2}} + v_{\mathrm{l_3}}
  = \sum\limits_{j=1}^{10} a_j f_j^{\mathrm{l}}(R,l,b)
  \nonumber \\
  v_{\mathrm{b}} = v_{\mathrm{b_1}} + v_{\mathrm{b_2}} + v_{\mathrm{b_3}}
  = \sum\limits_{j=1}^{10} a_j f_j^{\mathrm{b}}(R,l,b)
\end{eqnarray}

\noindent where the constants $a_j$ contain combinations of the kinematic
parameters that we wish to determine:
\begin{eqnarray}\label{eq.a}
a_1 & = & U_\odot \nonumber \\
a_2 & = & V_\odot \nonumber \\
a_3 & = & W_\odot \nonumber \\
a_4 & = & \Theta(\varpi_\odot) \nonumber \\
a_5 & = & a_{\mathrm{r}} \nonumber \\
a_6 & = & b_{\mathrm{r}} \nonumber \\
a_7 & = & \Pi_{\mathrm{b}} \cos \psi_\odot \nonumber \\
a_8 & = & \Pi_{\mathrm{b}} \sin \psi_\odot \nonumber \\
a_9 & = & \Theta_{\mathrm{b}} \sin \psi_\odot \nonumber \\
a_{10} & = & \Theta_{\mathrm{b}} \cos \psi_\odot
\end{eqnarray}

\noindent and $f_j^i(R,l,b)$ are functions of the heliocentric distance
and the galactic longitude and latitude:
\begin{eqnarray}\label{eq.f}
f_1^\mathrm{r} & = & - \cos l \cos b \nonumber \\
f_2^\mathrm{r} & = & - \sin l \cos b \nonumber \\
f_3^\mathrm{r} & = & - \sin b \nonumber \\
f_4^\mathrm{r} & = & \left[ \sin(l + \theta) - \sin l \right] \cos b
\nonumber \\
f_5^\mathrm{r} & = & \Delta \varpi \sin (l + \theta) \cos b \nonumber \\
f_6^\mathrm{r} & = & \Delta \varpi^2 \sin (l + \theta) \cos b \nonumber \\
\nonumber \\
f_7^\mathrm{r} & = & - \left( \cos \left[ m \left( \theta - \frac {\ln \frac
{\varpi}{\varpi_\odot}} {\tan i} \right) \right] \cos (l + \theta) -
f_\odot \cos l \right) \cos b \nonumber \\
\nonumber \\
f_8^\mathrm{r} & = & - \sin \left[ m \left( \theta - \frac {\ln \frac
{\varpi}{\varpi_\odot}} {\tan i} \right) \right] \cos (l + \theta) \cos b
\nonumber \\
\nonumber \\
f_9^\mathrm{r} & = & - \left( \cos \left[ m \left( \theta - \frac {\ln \frac
{\varpi}{\varpi_\odot}} {\tan i} \right) \right] \sin (l + \theta) -
f_\odot \sin l \right) \cos b \nonumber \\
\nonumber \\
f_{10}^\mathrm{r} & = & \sin \left[ m \left( \theta - \frac {\ln \frac
{\varpi}{\varpi_\odot}} {\tan i} \right) \right] \sin (l + \theta) \cos b
\nonumber \\
\nonumber \\
f_1^\mathrm{l} & = & \sin l \nonumber \\
f_2^\mathrm{l} & = & - \cos l \nonumber \\
f_3^\mathrm{l} & = & 0 \nonumber \\
f_4^\mathrm{l} & = & \cos(l + \theta) - \cos l \nonumber \\
f_5^\mathrm{l} & = & \Delta \varpi \cos(l + \theta) \nonumber \\
f_6^\mathrm{l} & = & \Delta \varpi^2 \cos(l + \theta) \nonumber \\
\nonumber \\
f_7^\mathrm{l} & = & \cos \left[ m \left( \theta - \frac {\ln \frac
{\varpi}{\varpi_\odot}} {\tan i} \right) \right] \sin(l + \theta) -
f_\odot \sin l \nonumber \\
\nonumber \\
f_8^\mathrm{l} & = & \sin \left[ m \left( \theta - \frac {\ln \frac
{\varpi}{\varpi_\odot}} {\tan i} \right) \right] \sin(l + \theta)
\nonumber \\
\nonumber \\
f_9^\mathrm{l} & = & - \cos \left[ m \left( \theta - \frac {\ln \frac
{\varpi}{\varpi_\odot}} {\tan i} \right) \right] \cos(l + \theta) -
f_\odot \cos l \nonumber \\
\nonumber \\
f_{10}^\mathrm{l} & = & \sin \left[ m \left( \theta - \frac {\ln \frac
{\varpi}{\varpi_\odot}} {\tan i} \right) \right] \cos(l + \theta)
\nonumber \\
\nonumber \\
f_1^\mathrm{b} & = & \cos l \sin b \nonumber \\
f_2^\mathrm{b} & = & \sin l \sin b \nonumber \\
f_3^\mathrm{b} & = & - \cos b \nonumber \\
f_4^\mathrm{b} & = & - \left[ \sin(l + \theta) - \sin l \right] \sin b
\nonumber \\
f_5^\mathrm{b} & = & - \Delta \varpi \sin (l + \theta) \sin b \nonumber \\
f_6^\mathrm{b} & = & - \Delta \varpi^2 \sin (l + \theta) \sin b \nonumber \\
\nonumber \\
f_7^\mathrm{b} & = & \left( \cos \left[ m \left( \theta - \frac {\ln \frac
{\varpi}{\varpi_\odot}} {\tan i} \right) \right] \cos (l + \theta) -
f_\odot \cos l \right) \sin b \nonumber \\
\nonumber \\
f_8^\mathrm{b} & = & \sin \left[ m \left( \theta - \frac {\ln \frac
{\varpi}{\varpi_\odot}} {\tan i} \right) \right] \cos (l + \theta) \sin b
\nonumber \\
\nonumber \\
f_9^\mathrm{b} & = & \left( \cos \left[ m \left( \theta - \frac {\ln \frac
{\varpi}{\varpi_\odot}} {\tan i} \right) \right] \sin (l + \theta) -
f_\odot \sin l \right) \sin b \nonumber \\
\nonumber \\
f_{10}^\mathrm{b} & = & - \sin \left[ m \left( \theta - \frac {\ln \frac
{\varpi}{\varpi_\odot}} {\tan i} \right) \right] \sin (l + \theta) \sin b
\end{eqnarray}

In our resolution procedure the parameters $a_j$ are computed following an
iterative scheme (extensively explained in Fern\'andez \cite{Fernandez})
and, from these, the kinematic parameters $U_\odot$, $V_\odot$, $W_\odot$,
$a_{\mathrm{r}}$, $b_{\mathrm{r}}$, $\psi_\odot$, $\Pi_{\mathrm{b}}$,
$\Theta_{\mathrm{b}}$ and $f_\odot$ are derived.

With regard to the free parameters of our model, different values for $m$,
$i$, $\varpi_\odot$, $\Theta(\varpi_\odot)$ were considered (see Sect.
\ref{discussion}).

%
\section{simulations to check the kinematic analysis}\label{ApenB}

In Paper I we did numerical simulations in order to evaluate the biases in
the kinematic model parameters (in that case, the Oort constants and the
solar motion components) induced by our observational constraints and
errors. In the present work, we have also generated simulated samples in
the same way, though the significant correlations detected between some
parameters make it advisable, in this case, to carry out a more detailed
study.

In this section we present the process used to generate the simulated
samples (the same as in Paper I, except for the change in the systematic
contributions considered), the results we obtained and, finally, the
quantification of the biases present in our real samples.

\subsection{Process used to generate the simulated samples}

To take into account the irregular spatial distribution of our stars and
their observational errors, parameters describing the position of each
simulated pseudo-star were generated as follow:
  
\begin{itemize}
            
\item From each real star we generated a pseudo-star that has the same
nominal position $(R_0,l,b)$ --not affected by errors-- as the real one.

\item We assumed that the angular coordinates ($l,b$) have negligible
observational errors.

\item The distance error of the pseudo-star has a distribution law:
\begin{equation}\label{eq.dist.fot}
  \varepsilon(R) = e^{
  - \frac {1}{2} \left( \frac {R - R_0}{\sigma_R} \right)^2}   
\end{equation}

\noindent where $\sigma_R$ is the individual error in the photometric
distance of the real star ($R_0$).

\end{itemize}

To generate the kinematic parameters we randomly assigned to each
pseudo-star a velocity $(U,V,W)$ by assuming a cosmic dispersion
$(\sigma_U, \sigma_V, \sigma_W)$ and a Schwarzschild distribution:
\begin{equation}
  \varphi'_v(U, V, W) = e^{
  - \frac {1}{2} \left( \frac {U - U_0}{\sigma_U} \right)^2
  - \frac {1}{2} \left( \frac {V - V_0}{\sigma_V} \right)^2
  - \frac {1}{2} \left( \frac {W - W_0}{\sigma_W} \right)^2}
\end{equation}

\noindent where $(U_0,V_0,W_0)$ are the reflex of solar motion. These
components were transformed into radial velocities and proper motions in
galactic coordinates using the nominal position of the pseudo-star
$(R_0,l,b)$. The systematic motion due to galactic rotation and spiral arm
kinematics was added following Eqs. (\ref{eq.v2}) and (\ref{eq.v3}),
obtaining the components $(v_{\mathrm{r_0}}, \mu_{\mathrm{l_0}},
\mu_{\mathrm{b_0}})$ for each pseudo-star. Finally, individual
observational errors were introduced by using the error function:
\begin{equation}
  \varepsilon(v_{\mathrm{r}}, \mu_{\mathrm{l}}, \mu_{\mathrm{b}}) = e^{
  - \frac {1}{2} \left( \frac {v_{\mathrm{r}} - v_{\mathrm{r_0}}}
  {\sigma_{v_{\mathrm{r}}}} \right)^2
  - \frac {1}{2} \left( \frac {\mu_{\mathrm{l}} - \mu_{\mathrm{l_0}}}
    {\sigma_{\mu_{\mathrm{l}}}} \right)^2
  - \frac {1}{2} \left( \frac {\mu_{\mathrm{b}} - \mu_{\mathrm{b_0}}}
    {\sigma_{\mu_{\mathrm{b}}}} \right)^2}
\end{equation}

\noindent where $\sigma_{v_{\mathrm{r}}}$, $\sigma_{\mu_{\mathrm{l}}}$ and
$\sigma_{\mu_{\mathrm{b}}}$ are the observational errors of the real star.

At the end of this process we had the following data for each pseudo-star:
galactic coordinates ($R,l,b$), velocity parameters ($v_{\mathrm{r}},
\mu_{\mathrm{l}}, \mu_{\mathrm{b}}$), errors in the velocity parameters
($\sigma_{v_{\mathrm{r}}}, \sigma_{\mu_{\mathrm{l}}},
\sigma_{\mu_{\mathrm{b}}}$) and error in the photometric distance
($\sigma_R$). The simulated radial component of those pseudo-stars
generated from a real star without radial velocity was not used, thus we
imposed on the simulated sample the same deficiency in radial velocity
data that is present in our real sample (see Sect. 2.2 and Appendix B in
Paper I for more details).

Following this scheme, several sets of 50 simulated samples for both O and
B stars and Cepheids were built. A classical solar motion of $(U,V,W) =
(9,12,7)$ km s$^{-1}$ was considered, taking the dispersion velocity
components $(\sigma_U,\sigma_V,\sigma_W) = (8,8,5)$ km s$^{-1}$ for O and
B stars (see Paper I) and $(\sigma_U,\sigma_V,\sigma_W) = (13,13,6)$ km
s$^{-1}$ for Cepheids (Luri \cite{Luri}). For the galactic rotation
parameters, we chose the values $a_{\mathrm{r}} = -2.1$ km s$^{-1}$
kpc$^{-1}$ and $b_{\mathrm{r}} = 0.0$ km s$^{-1}$ kpc$^{-2}$, which
correspond to a linear rotation curve with an $A$ Oort constant of 14.0 km
s$^{-1}$ kpc$^{-1}$. On the other hand, for the spiral structure
parameters several sets of values were used for $\psi_\odot$ (from
$\psi_\odot = 0\degr$ to $\psi_\odot = 315\degr$, in steps of $45\degr$)
and $\Omega_{\mathrm{p}}$ (from $\Omega_{\mathrm{p}} = 10$ km s$^{-1}$
kpc$^{-1}$ to $\Omega_{\mathrm{p}} = 40$ km s$^{-1}$ kpc$^{-1}$, in steps
of 5 km s$^{-1}$ kpc$^{-1}$), whereas a fixed value of $f_{\mathrm{r}} =
0.05$ was considered (Yuan \cite{Yuan}). From
$(\sigma_U,\sigma_V,\sigma_W)$, $a_{\mathrm{r}}$, $\Omega_{\mathrm{p}}$
and $f_{\mathrm{r}}$, the values of $\Pi_{\mathrm{b}}$,
$\Theta_{\mathrm{b}}$ and $f_\odot$ were inferred for each set of samples.

56 sets of 50 samples for both O and B stars and Cepheids were generated.
Concerning the free parameters of our model, in a first stage we adopted
classical values ($m = 2$, $i = -6\degr$, Lin et al. \cite{Lin et al.2};
$\varpi_\odot = 8.5$ kpc, $\Theta(\varpi_\odot) =220$ km s$^{-1}$, Kerr \&
Lyndell-Bell \cite{Kerr et al.}), though we also tested cases with $m =
4$, $i = -12\degr$ (Amaral \& L\'epine \cite{Amaral et al.}) and
$\varpi_\odot = 7.1$ kpc, $\Theta(\varpi_\odot) =184$ km s$^{-1}$ (Olling
\& Merrifield \cite{Olling et al.}). In Table \ref{tab.simpar} we
summarize all the adopted kinematic parameters.

\begin{table}
\caption[]{Parameters of the simulated samples.}
\label{tab.simpar}
\begin{tabular}{ll}
        \hline
$U_\odot$             &  9 km \mbox{s$^{-1}$} \\
$V_\odot$             & 12 km \mbox{s$^{-1}$} \\
$W_\odot$             &  7 km \mbox{s$^{-1}$} \\
\hline
$(\sigma_U, \sigma_V, \sigma_W)$ & $(8,8,5)$ km s$^{-1}$ (O and B stars) \\
                                 & $(13,13,6)$ km s$^{-1}$ (Cepheid stars) \\
\hline
$a_\mathrm{r}$        & $-$2.1 km \mbox{s$^{-1}$} \mbox{kpc$^{-1}$} \\
$b_\mathrm{r}$        &    0.0 km \mbox{s$^{-1}$} \mbox{kpc$^{-2}$} \\
\hline
$f_{\mathrm{r}}$      &    0.05 \\
$\psi_\odot$          & from 0$\degr$ to 360$\degr$, \\
                      & in steps of 45$\degr$ \\
$\Omega_{\mathrm{p}}$ & from 10 km s$^{-1}$ kpc$^{-1}$ to 40 km s$^{-1}$ kpc$^{-1}$, \\
                      & in steps of 5 km s$^{-1}$ kpc$^{-1}$ \\
\hline
Case A                & $m = 2$, $i = -6\degr$, \\
                      & $\varpi_\odot =8.5$ kpc, $\Theta(\varpi_\odot) = 220$ km s$^{-1}$ \\
Case B                & $m = 2$, $i = -6\degr$, \\
                      & $\varpi_\odot =7.1$ kpc, $\Theta(\varpi_\odot) = 184$ km s$^{-1}$ \\
Case C                & $m = 4$, $i = -12\degr$, \\
                      & $\varpi_\odot =8.5$ kpc, $\Theta(\varpi_\odot) = 220$ km s$^{-1}$ \\
Case D                & $m = 4$, $i = -12\degr$, \\
                      & $\varpi_\odot =7.1$ kpc, $\Theta(\varpi_\odot) = 184$ km s$^{-1}$ \\
\hline
\end{tabular}
\end{table}

\subsection{Results and discussion}

\subsubsection{Results for a 2-armed model of the Galaxy}

\begin{figure}
  \resizebox{8.5cm}{!}{\includegraphics{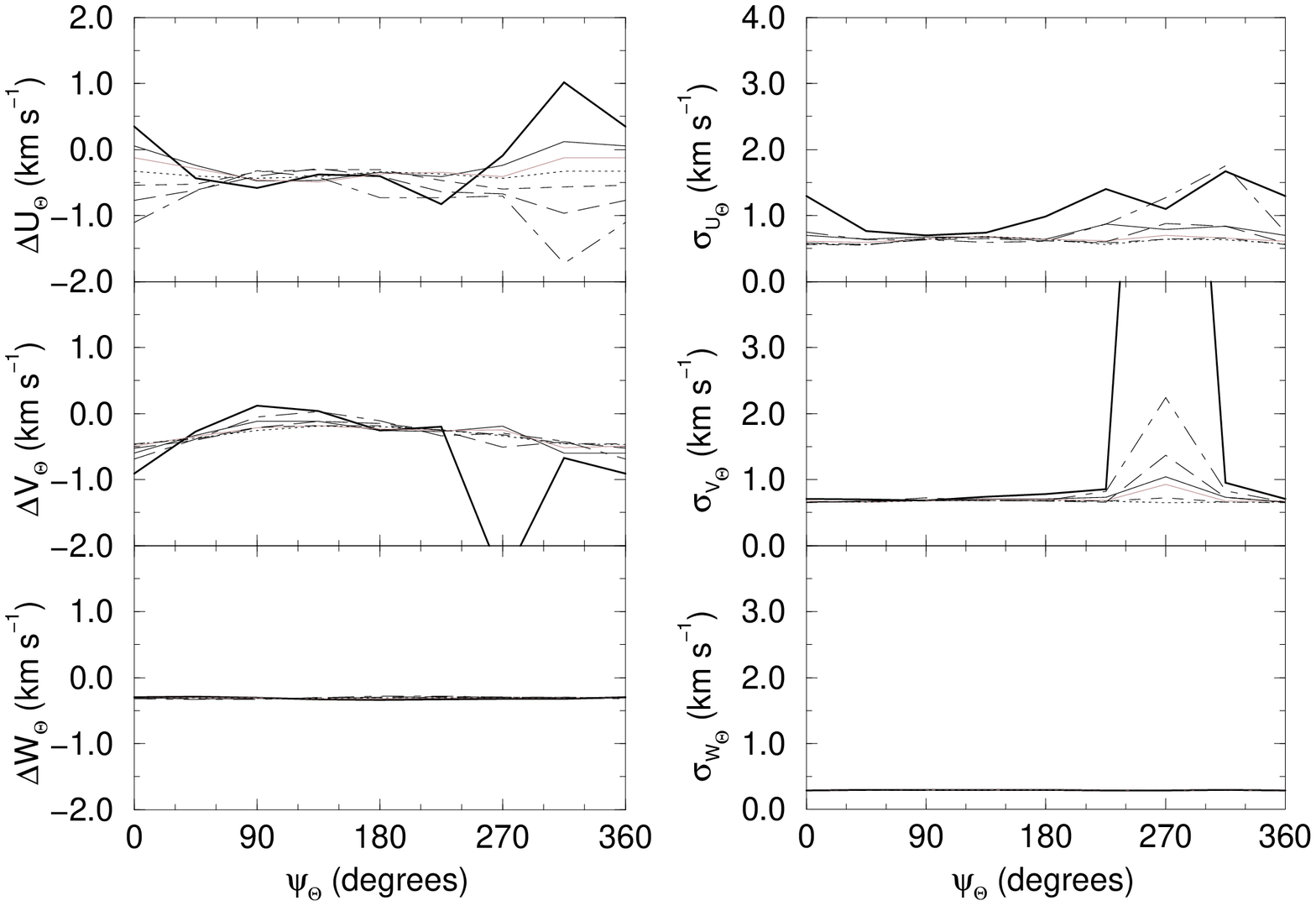}}
  \resizebox{8.5cm}{!}{\includegraphics{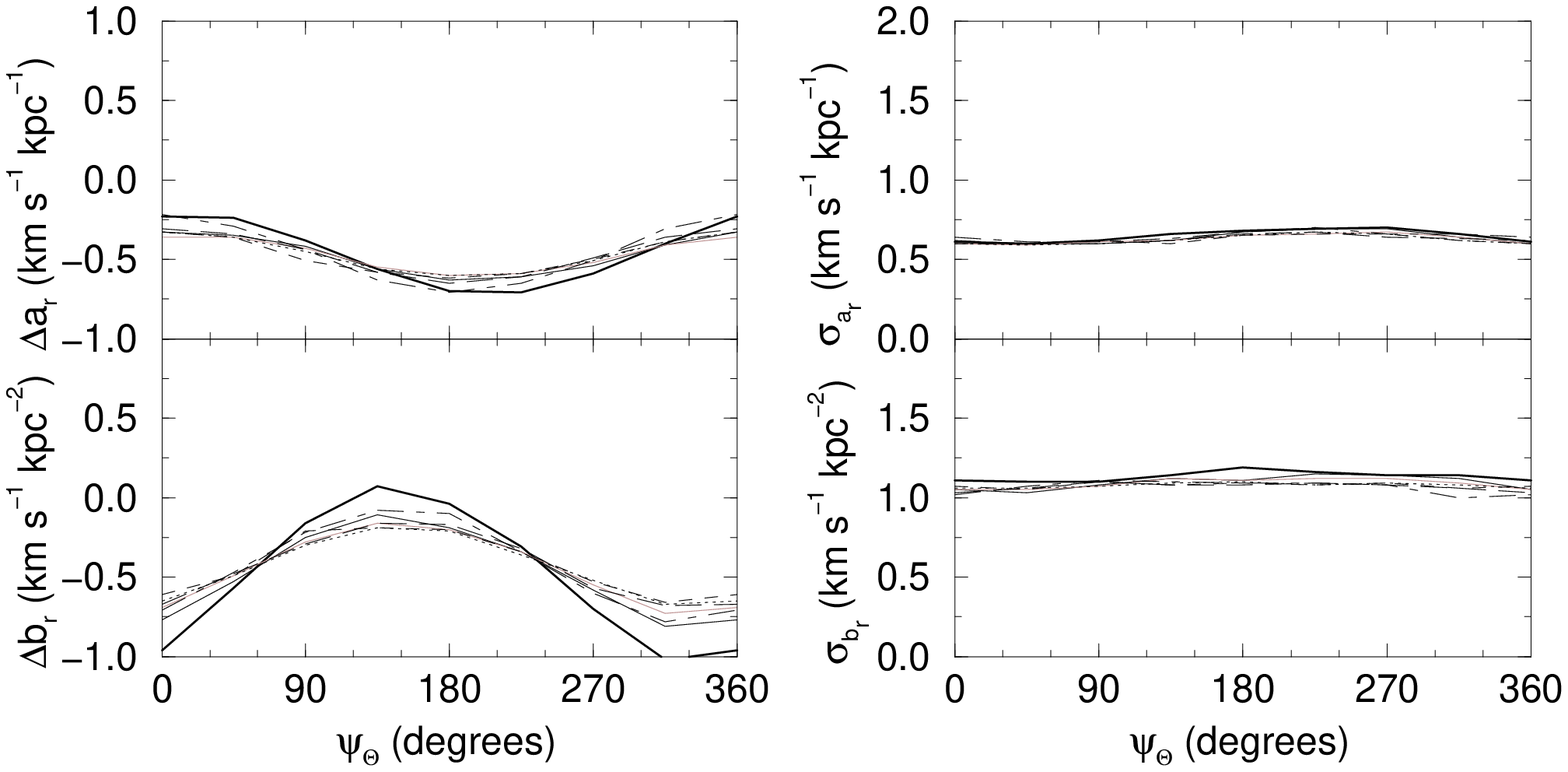}}
  \resizebox{8.5cm}{!}{\includegraphics{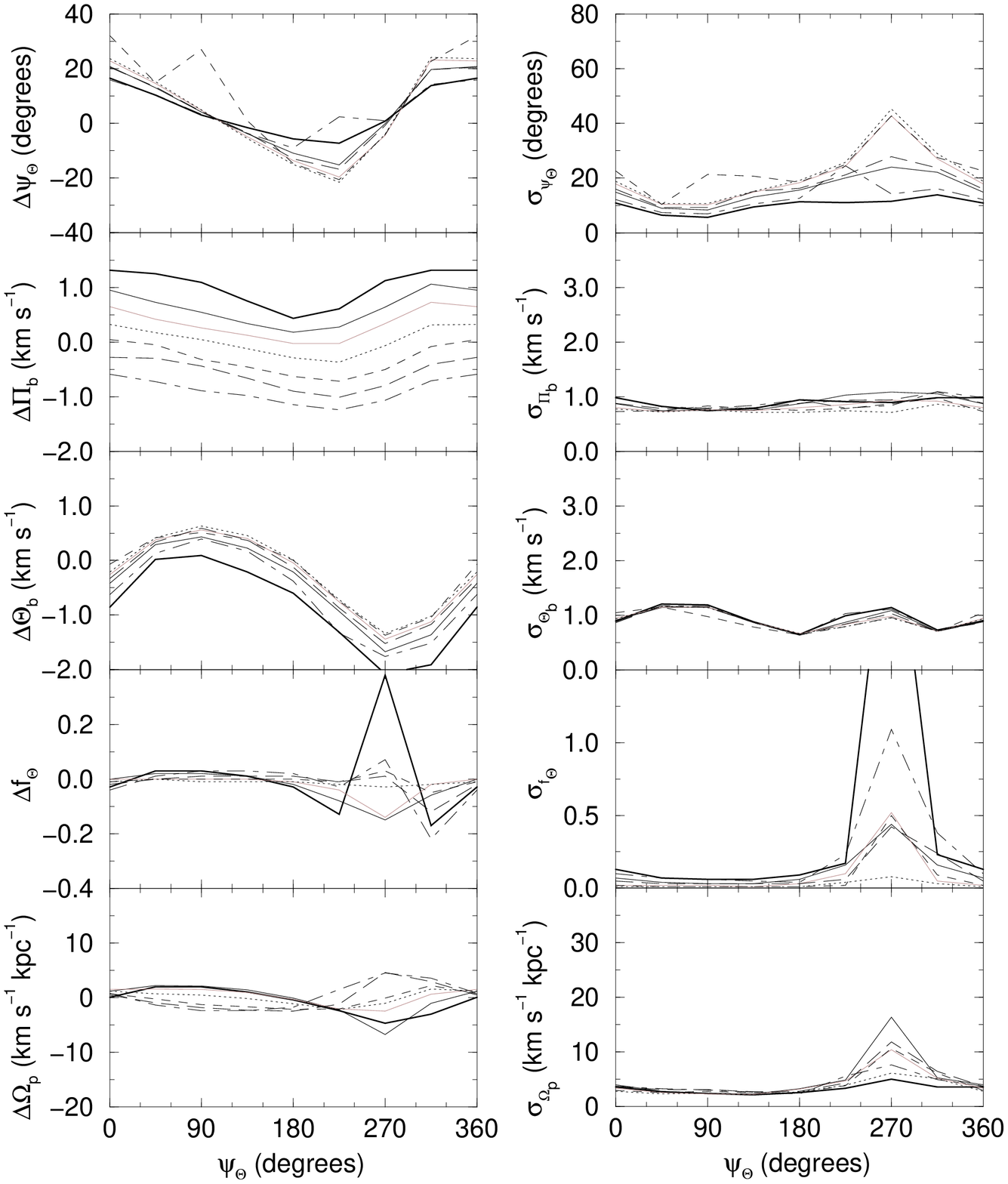}}
  \caption{Bias (obtained value$-$simulated value; left) and standard
           deviation (right) for each set of 50 samples in the solar
           motion (top), galactic rotation (middle) and spiral arm
           kinematic (bottom) parameters for O and B pseudo-stars (Case
           A). Values of $\Omega_{\mathrm{p}}$: 10 km s$^{-1}$ kpc$^{-1}$
           (black double solid line), 15 (black solid line), 20 (grey
           solid line), 25 (dotted line), 30 (dashed line), 35 (long
           dashed line) and 40 (dot-dashed line).}
  \label{fig.OBsim}
\end{figure}

\begin{figure}
  \resizebox{8.5cm}{!}{\includegraphics{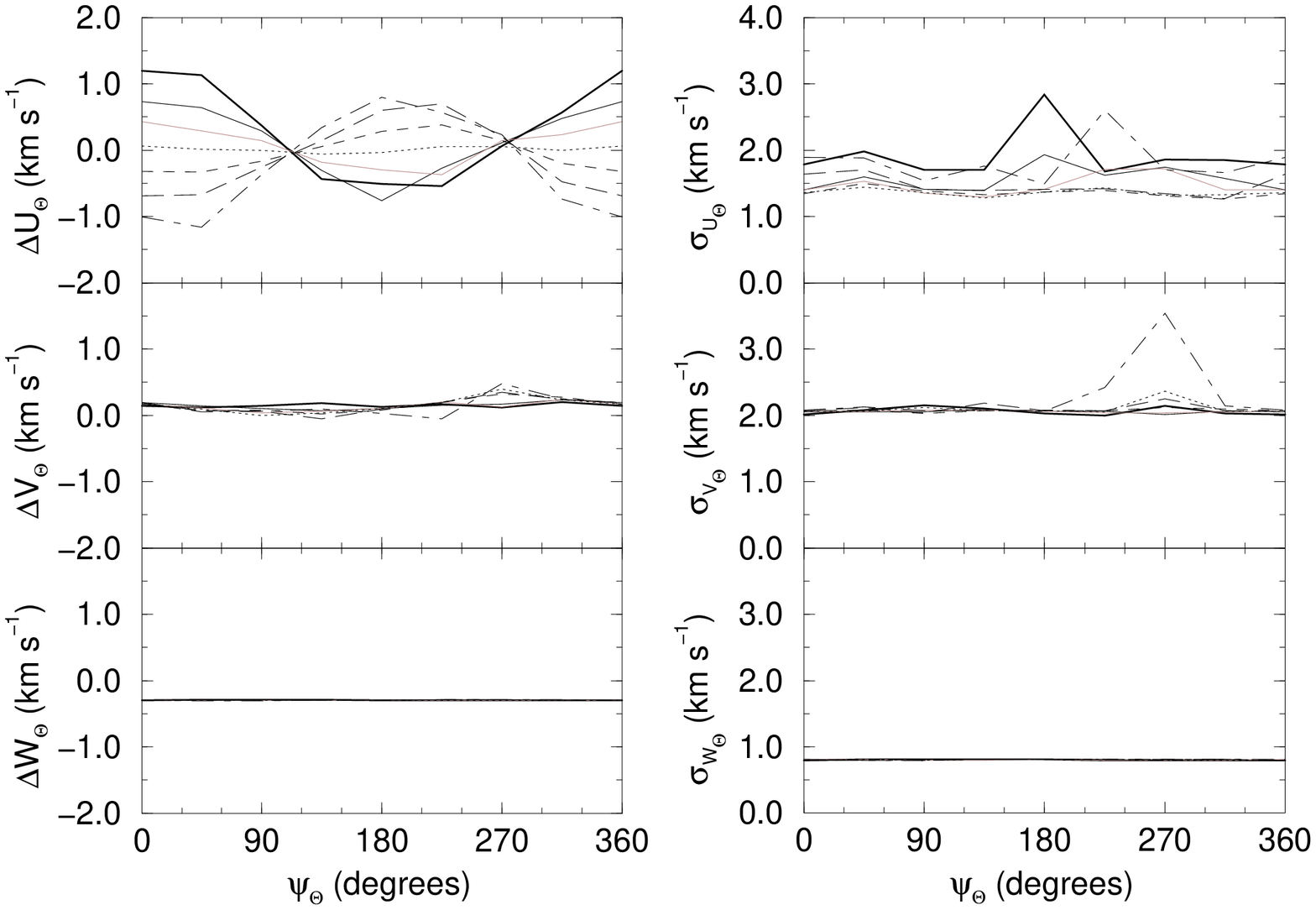}}
  \resizebox{8.5cm}{!}{\includegraphics{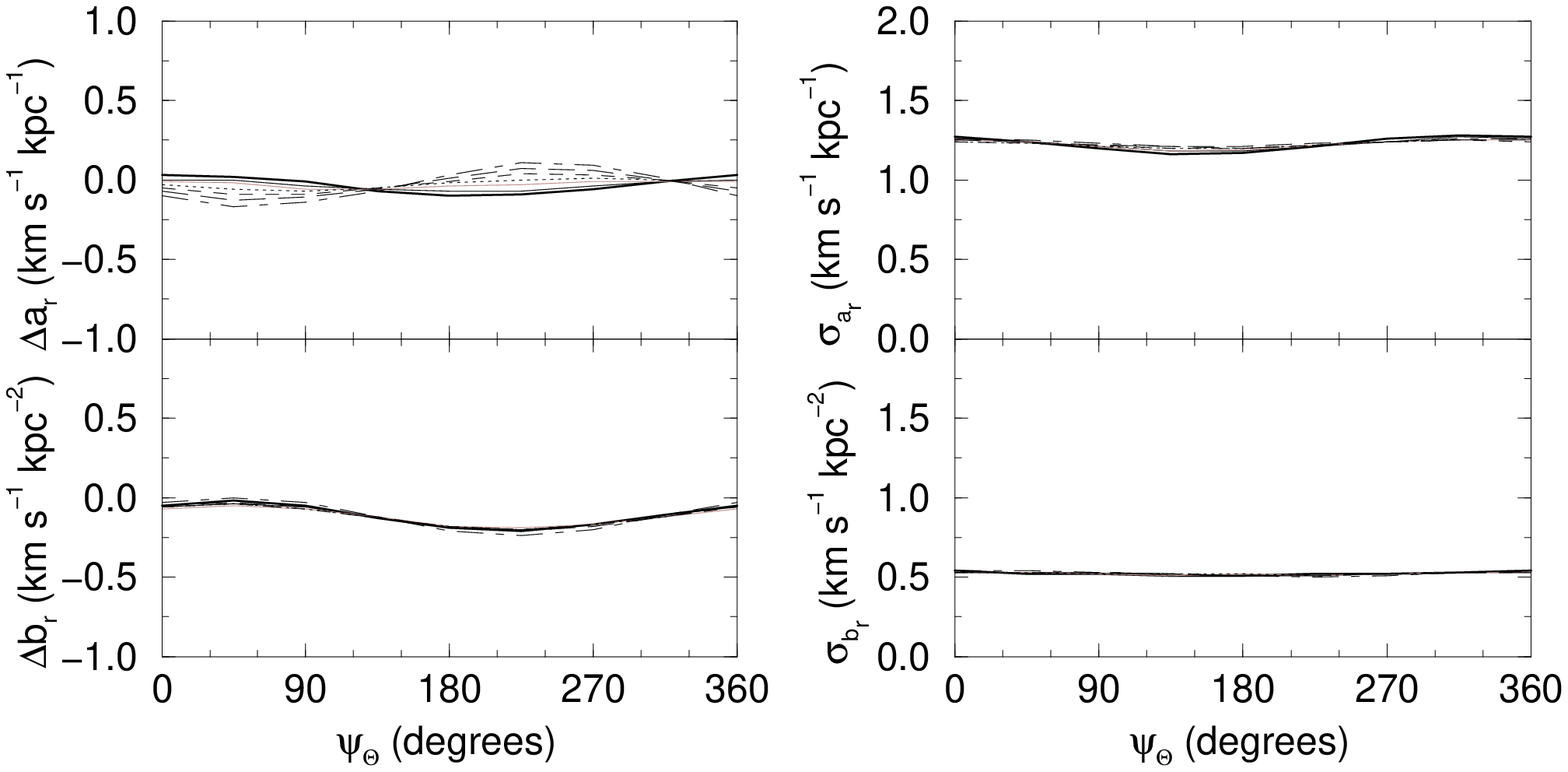}}
  \resizebox{8.5cm}{!}{\includegraphics{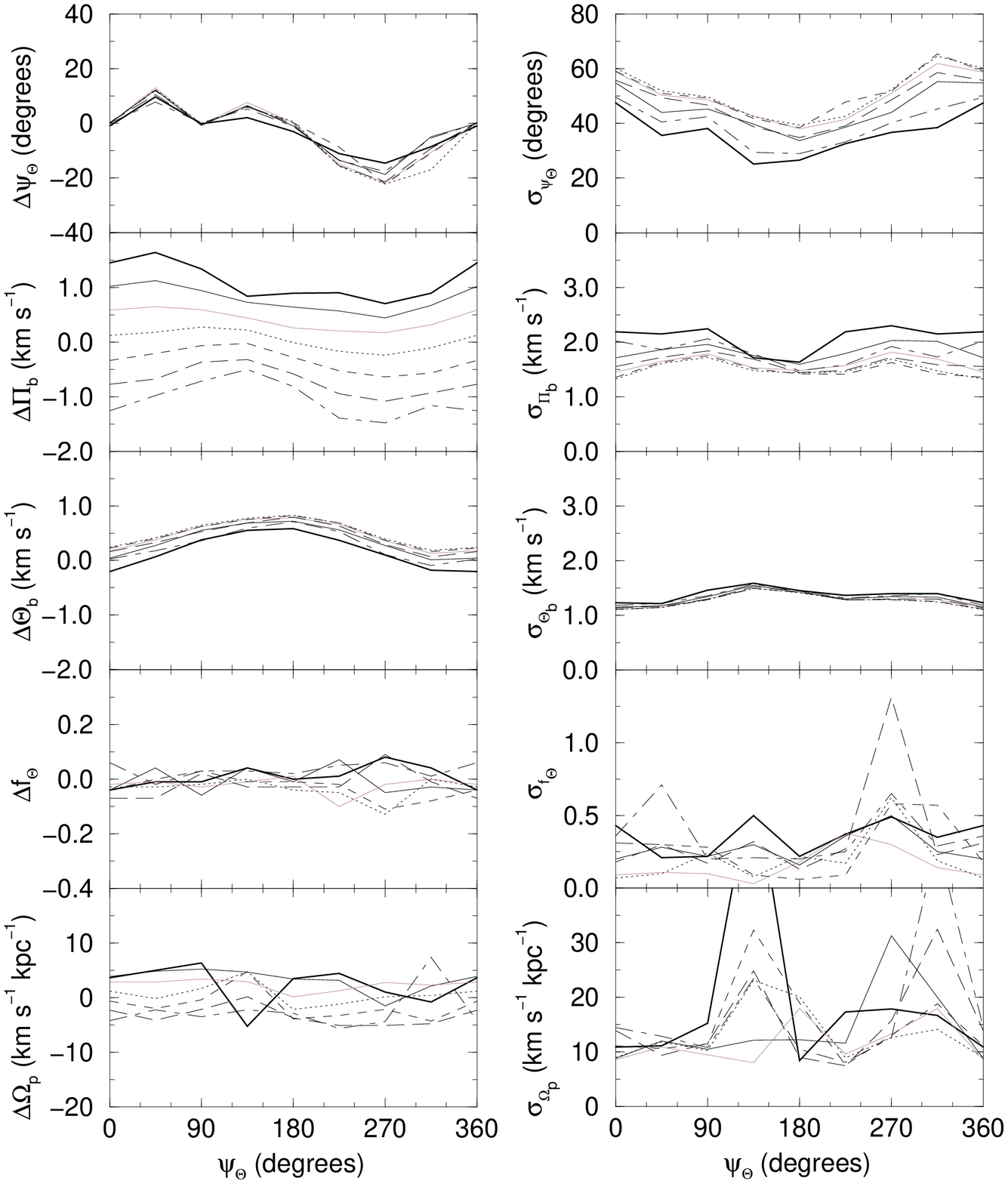}}
  \caption{Bias (obtained value$-$simulated value; left) and standard
           deviation (right) for each set of 50 samples in the solar
           motion (top), galactic rotation (middle) and spiral arm
           kinematic (bottom) parameters for  Cepheid pseudo-stars (Case 
           A). Values of $\Omega_{\mathrm{p}}$: 10 km s$^{-1}$ kpc$^{-1}$
           (black double solid line), 15 (black solid line), 20 (grey
           solid line), 25 (dotted line), 30 (dashed line), 35 (long
           dashed line) and 40 (dot-dashed line).}
  \label{fig.Cefsim}
\end{figure}

A complete solution simultaneously taking into account radial velocity and
proper motion data was computed. Our test showed that the number of
Cepheids within 2 kpc from the Sun prevents the obtainment of reliable
results.  In Figs. \ref{fig.OBsim} and \ref{fig.Cefsim} we show the
results obtained for the simulated samples of O and B stars (0.6 $< R <$
2 kpc) and Cepheids (0.6 $< R <$ 4 kpc) in case A (see Table
\ref{tab.simpar}).

As a first conclusion, and confirming our suspicions, a systematic trend
with $\psi_\odot$ and/or $\Omega_{\mathrm{p}}$ is observed in most cases.
This behaviour is produced by the correlations between some terms in the
least squares fit, which depends on the spatial distribution of each
sample.

For solar motion a bias between $-1.5$ and $1.5$ km s$^{-1}$ (depending on
$\psi_\odot$ and $\Omega_{\mathrm{p}}$) was found for $U_\odot$ and
$V_\odot$, and of only $-0.3$ km s$^{-1}$ for $W_\odot$. For both O and B
stars and Cepheids we found the bias on $V_\odot$ and $W_\odot$ to be
independent of $\Omega_{\mathrm{p}}$, with a slight dependence on
$\psi_\odot$. For $\psi_\odot = 270\degr$ and $\Omega_{\mathrm{p}} = 10$
km s$^{-1}$ kpc$^{-1}$ we found a large negative bias on $V_\odot$, but
with a great standard deviation. This occurred in several samples (inside
this set) with serious convergence problems in the iteration procedure we
use to solve the least squares fit. Similar problems in other cases with
$\psi_\odot = 270\degr$ will be found later. For O and B stars, the
standard deviations in the solar motion components are $\approx 0.6$ km
s$^{-1}$ for $U_\odot$ and $V_\odot$ (except for $\psi_\odot = 270\degr$),
and $\approx 0.3$ km s$^{-1}$ for $W_\odot$. On the other hand, in the
case of Cepheids these values increased to $\approx$ 1.4-2.0 km s$^{-1}$
and $\approx 0.8$ km s$^{-1}$, respectively.

The biases found in the first- and second-order terms of the galactic
rotation curve are negligible for Cepheids, with a level fluctuation of
$\pm 0.3$ km s$^{-1}$ kpc$^{-1}$ (or km s$^{-1}$ kpc$^{-2}$). In this case
there is a standard deviation of 1.3 for $a_{\mathrm{r}}$ and 0.5 for
$b_{\mathrm{r}}$. For O and B stars the biases clearly depend on
$\psi_\odot$, varying from $-0.7$ to $-0.5$ km s$^{-1}$ kpc$^{-1}$ for
$a_{\mathrm{r}}$, and from $-1.0$ to $0.1$ km s$^{-1}$ kpc$^{-2}$ for
$b_{\mathrm{r}}$. The standard deviations are 0.8 km s$^{-1}$ kpc$^{-1}$
and 1.2 km s$^{-1}$ kpc$^{-2}$, respectively.

Let us study the biases that have an effect on the determination of the
spiral structure parameters. As a general conclusion, Figs.
\ref{fig.OBsim} and \ref{fig.Cefsim} show that our sample of O and B stars
supplies better results than the Cepheid sample.

In the case of O and B stars, we found a clear dependence with
$\psi_\odot$ and $\Omega_\mathrm{p}$ in $\psi_\odot$, $\Pi_\mathrm{b}$ and
$\Theta_\mathrm{b}$ determinations, whereas $f_\odot$ and
$\Omega_\mathrm{p}$ only show peculiar behaviour around $\psi_\odot =
270\degr$. Concerning $\psi_\odot$, the bias oscillates from $-20\degr$ to
$30\degr$. The standard deviation of the mean for the 50 samples of each
set is about 10-$20\degr$. On the other hand, for the amplitudes
$\Pi_\mathrm{b}$ and $\Theta_\mathrm{b}$ the biases are of $\pm 2$ km
s$^{-1}$, with a standard deviation of about 1 km s$^{-1}$. Neither
$f_\odot$ nor $\Omega_\mathrm{p}$ have a considerable bias, except for
$\psi_\odot = 270\degr$, where both biases and standard deviations go up.

For Cepheid stars similar results were obtained, but with larger standard
deviations in all cases. The bias in $\psi_\odot$ changes from $-20\degr$
to $10\degr$, with standard deviations of about 30-$60\degr$. In the case
of $\Pi_\mathrm{b}$, we found a clear dependence on both $\psi_\odot$ and
$\Omega_\mathrm{p}$, with a bias of $\pm1.5$ km s$^{-1}$ and a standard
deviation of about 2 km s$^{-1}$. On the other hand, for
$\Theta_\mathrm{b}$ the bias is smaller, from 0 to 0.8 km s$^{-1}$, and
the standard deviation is 1-1.5 km s$^{-1}$. As for O and B stars, for
$f_\odot$ and $\Omega_\mathrm{p}$ small biases are found, though the
standard deviations are larger in this case.

\subsubsection{Results considering possible errors in the choice of the
free parameters}

An interesting point to analyse is the study of the biases produced by a
bad choice of the free parameters in our model ($m$, $i$, $\varpi_\odot$,
$\Theta(\varpi_\odot)$). In the same way as in the previous section, we
simulated 50 samples for each one of the cases considered in the real
resolution, i.e. cases A, B, C and D (see Table \ref{tab.sim}). The
simulated parameters were the same as in Table \ref{tab.simpar} for solar
motion and galactic rotation. For spiral arm kinematics, we considered
$\psi_\odot = 315\degr$ and $\Omega_{\mathrm{p}} = 30$ km s$^{-1}$
kpc$^{-1}$ (similar values to those obtained from real samples; see Sect.
\ref{results}).

\begin{table*}
\caption[]{Bias and standard deviation in $\psi_\odot$ and
           $\Omega_{\mathrm{p}}$ for crossed solutions for the simulated
           samples of O and B stars and Cepheids. Units: $\psi_\odot$ in
           degrees; $\Omega_{\mathrm{p}}$ in km s$^{-1}$ kpc$^{-1}$.}
\label{tab.sim}
\begin{tabular}{crrrrrrrr}
        \hline
        & \multicolumn{4}{c}{O and B stars with 0.6 $< R <$ 2 kpc}
        & \multicolumn{4}{c}{Cepheid stars with 0.6 $< R <$ 4 kpc} \\
        \hline
        \hline
        & Case A & Case B & Case C & Case D
        & Case A & Case B & Case C & Case D \\
        \hline
        \hline
        & \multicolumn{8}{c}{Case A simulated} \\
        \hline
$\Delta \psi_\odot$            &   23.  &   17.  &   24.  &   19.
                               &$-$11.  &    0.  & $-$2.  &    3.  \\
$\sigma_{\psi_\odot}$          &   28.  &   27.  &   25.  &   25.  
                               &   65.  &   61.  &   71.  &   71.  \\
$\Delta \Omega_{\mathrm{p}}$   &    2.2 &    3.1 & $-$1.1 & $-$0.8 
                               & $-$4.3 & $-$3.5 & $-$2.5 & $-$4.0 \\
$\sigma_{\Omega_{\mathrm{p}}}$ &    6.2 &    7.3 &    2.9 &    3.5 
                               &   18.9 &   16.6 &    8.8 &    7.3 \\
        \hline
        & \multicolumn{8}{c}{Case B simulated} \\
        \hline
$\Delta \psi_\odot$            &   32.  &   24.  &   32.  &   25.  
                               &    0.  &    4.  &$-$11.  &    1.  \\
$\sigma_{\psi_\odot}$          &   34.  &   32.  &   31.  &   30.  
                               &   83.  &   75.  &   82.  &   78.  \\
$\Delta \Omega_{\mathrm{p}}$   &    0.7 &    1.9 & $-$1.6 & $-$1.3 
                               & $-$4.3 &$-$14.3 & $-$4.3 & $-$4.1 \\
$\sigma_{\Omega_{\mathrm{p}}}$ &    7.1 &    8.0 &    3.8 &    3.9 
                               &   16.2 &   52.8 &    7.0 &    8.1 \\
        \hline
        & \multicolumn{8}{c}{Case C simulated} \\
        \hline
$\Delta \psi_\odot$            &   18.  &   14.  &   17.  &   14.  
                               &    4.  &    2.  & $-$2.  &   10.  \\
$\sigma_{\psi_\odot}$          &   24.  &   23.  &   21.  &   21.  
                               &   62.  &   58.  &   60.  &   63.  \\
$\Delta \Omega_{\mathrm{p}}$   &    6.7 &    8.2 &    1.2 &    2.0 
                               & $-$2.4 & $-$1.1 & $-$1.0 & $-$3.0 \\
$\sigma_{\Omega_{\mathrm{p}}}$ &    6.5 &    7.3 &    2.9 &    3.3 
                               &   25.1 &   21.5 &   10.9 &    9.2 \\
        \hline
        & \multicolumn{8}{c}{Case D simulated} \\
        \hline
$\Delta \psi_\odot$            &   29.  &   23.  &   25.  &   21.  
                               &    9.  &   10.  & $-$9.  &    1.  \\
$\sigma_{\psi_\odot}$          &   31.  &   29.  &   27.  &   26.  
                               &   84.  &   78.  &   81.  &   73.  \\
$\Delta \Omega_{\mathrm{p}}$   &    4.3 &    6.3 &    0.2 &    1.1 
                               & $-$3.9 & $-$4.9 & $-$3.2 & $-$2.8 \\
$\sigma_{\Omega_{\mathrm{p}}}$ &    7.1 &    8.2 &    3.7 &    3.8 
                               &   17.1 &   19.5 &    7.8 &    8.6 \\
        \hline
        \hline
\end{tabular}
\end{table*}

In Table \ref{tab.sim} we show the biases and standard deviations when
solving the model equations in crossed solutions (e.g. we generated 50
simulated samples considering the free parameters in case A, and then we
solved equations using the free parameters adopted for cases A, B, C and
D, and so on for the other cases). As a first conclusion, we can observe
that a bad choice in the free parameters does not substantially alter the
derived kinematic parameters, particularly $\psi_\odot$. In other words,
for each set of simulated samples we obtained nearly the same values for
the parameters whether we solved the Eqs. (\ref{eq.fit}) with the correct
set of free parameters or with a wrong combination of them. Differences in
$\psi_\odot$ do not exceed 10$\degr$ for O and B stars and 20$\degr$ for
Cepheids. In the case of $\Omega_{\mathrm{p}}$ we found large differences
in some cases, but always when the standard deviation was also large. This
is especially true for Cepheids. A remarkable point is that the minimum
bias was not always produced when we properly chose the free parameters.

\subsubsection{Conclusions}

In the light of these results, we conclude that we are able to determine
the kinematic parameters of the proposed model of the Galaxy from the real
star samples described in Sect. \ref{samples}, supposing that the velocity
field of the stars is well described by this model. We studied case A ($m
= 2$, $i = -6\degr$, $\varpi_\odot = 8.5$ kpc, $\Theta(\varpi_\odot) =
220$ km s$^{-1}$ in detail in these simulations, but we also looked at the
other combinations of the free parameters (cases B, C and D), with similar
conclusions. Nevertheless, the study of crossed solutions has shown that
it will be very difficult to decide between the several set of free
parameters discussed in Sect. \ref{results} (see also Table
\ref{tab.sim}), owing to the small differences obtained when changing the
free parameters in the condition equations.

%


\begin{thebibliography}{}
   \bibitem[1997]{Amaral et al.} Amaral, L.H., \& L\'epine, J.R.D. 1997, 
      MNRAS, 286, 885
   \bibitem[1989]{Avedisova} Avedisova, V.S. 1989, 
      Astrophysics (Tr. Astrofizika), 30, 83
   \bibitem[1981]{Bash} Bash, F.N. 1981,
      ApJ, 250, 551
   \bibitem[1991]{Battinelli} Battinelli, P. 1991,
      A\&A, 244, 69
   \bibitem[1999]{Beaulieu} Beaulieu, J.-P. 1999, private communication
   \bibitem[1958]{Bok} Bok, B.J. 1958,
      Observatory, 79, 58
   \bibitem[1974]{Bok et al.} Bok, B.J., \& Bok, P.F. 1974, 
      The Milky Way, Cambridge, Harvard University Press
   \bibitem[1971]{Burton} Burton, W.B. 1971, 
      A\&A, 10, 76
   \bibitem[1976]{Burton2} Burton, W.B. 1976, 
      ARA\&A, 14, 275
   \bibitem[1978]{Byl et al.} Byl, J., \& Ovenden, M.W. 1978, 
      ApJ, 225, 496
   \bibitem[1991]{Comeron et al.1} Comer\'on, F., \& Torra, J. 1991,
      A\&A, 241, 57
   \bibitem[1994]{Comeron et al.2} Comer\'on, F., Torra, J., \& G\'omez,
                                   A.E. 1994,
      A\&A, 286, 789
   \bibitem[1973]{Creze et al.} Cr\'ez\'e, M., \& Mennessier, M.O. 1973, 
      A\&A, 27, 281
   \bibitem[2000]{Drimmel} Drimmel, R. 2000,
      A\&A, 358, L13
   \bibitem[1985]{Elmegreen} Elmegreen, D.M. 1985, 
      The Milky Way galaxy, eds. H. van W\"orden, R.J. Allen, W.B. Burton,
      IAU Symposium 106, 255
   \bibitem[1999]{Englmaier et al.} Englmaier, P., \& Gerhard, O. 1999,
      MNRAS, 304, 512
   \bibitem[1997]{ESA2} ESA 1997,
      The Hipparcos Catalogue, ESA SP-1200
   \bibitem[1997]{Feast et al.0} Feast, M.W., \& Catchpole, R.M. 1997,
      MNRAS, 286, L1
   \bibitem[1997]{Feast et al.1} Feast, M.W., \& Whitelock, P.A. 1997,
      MNRAS, 291, 683
   \bibitem[1998]{Feast et al.2} Feast, M.W., Pont, F., \& Whitelock, P.A. 
                                 1998,
      MNRAS, 298, L43
   \bibitem[1998]{Fernandez} Fern\'andez, D. 1998,
      Degree of Physics (Master Thesis), Universitat de Barcelona, Spain
      (available in Spanish language from 
      http://www.am.ub.es/$\sim$dfernand)
   \bibitem[1995]{Fernie et al.} Fernie, J.D., Beattie, B., Evans, N.R.,
                                 \& Seager, S. 1995,
      IBVS, 4148
   \bibitem[1996]{Frink et al.} Frink, S., Fuchs, B., R\"oser, S., \&
                                Wielen, R. 1996,
      A\&A, 314, 430
   \bibitem[1992]{Garmany et al.} Garmany, C.D., \& Stencel, R.E. 1992,
      A\&AS, 94, 211
   \bibitem[1976]{Georgelin et al.} Georgelin, Y.M., \& Georgelin,
                                    Y.P. 1976,
      A\&A, 49, 57
   \bibitem[1977]{Gomez et al.} G\'omez, A., \& Mennessier, M.O. 1977,
      A\&A, 54, 113
   \bibitem[1998]{Glushkova et al.} Glushkova, E.V., Dambis, A.K.,
                                    Mel'nik, A.M., \& Rastorguev,
                                    A.S. 1998,
      A\&A, 329, 514
   \bibitem[1997]{Grenier} Grenier, S. 1997,
      private communication
   \bibitem[1998]{Hauck et al.} Hauck, B., \& Mermilliod, J.C. 1998, 
      A\&AS, 129, 431
   \bibitem[1969]{Kerr} Kerr, F.J. 1969
      ARA\&A, 7, 39
   \bibitem[1986]{Kerr et al.} Kerr, F.J., \& Lynden-Bell, D. 1986, 
      MNRAS, 221, 1023
   \bibitem[2001]{Lepine et al.} L\'epine, J.R.D., Mishurov, Yu.N., \&
                                 Dedikov, S.Yu. 2001,
      AJ, 546, 234
   \bibitem[1964]{Lin et al.1} Lin, C.C., \& Shu, F.H. 1964,
      ApJ, 140, 646
   \bibitem[1969]{Lin et al.2} Lin, C.C., Yuan, C., \& Shu, F.H. 1969,
      ApJ, 155, 721
   \bibitem[2000]{Luri} Luri, X., 2000, private communication
   \bibitem[1993]{Maciel} Maciel, W.J. 1993
      Ap\&SS, 206, 285
   \bibitem[1972]{Marochnik et al.} Marochnik, L.S., Mishurov, Yu.N.,
                                    \& Suchkov, A.A. 1972,
      Ap\&SS, 19, 285
   \bibitem[1995]{Mel'nik et al.1} Mel'nik, A.M., \& Efremov, Yu.N. 1995,
      Astro. Lett., 21, 10
   \bibitem[1998]{Mel'nik et al.2} Mel'nik, A.M., Sitnik, T.G., Dambis,
                                   A.K., Efremov, Yu.N., \& Rastorguev,
                                   A.S. 1998,
      Astro. Lett., 24, 594
   \bibitem[1999]{Mel'nik et al.3} Mel'nik, A.M., Dambis, A.K., \&
                                   Rastorguev, A.S. 1999,
      Astro. Lett., 25, 518
   \bibitem[1975]{Mennessier at al.} Mennessier, M.O., \& Cr\'ez\'e,
                                     M. 1975,
      in "La dynamique des galaxies spirales", colloque n$^\circ$241,
      Centre National de la Recherche Scientifique, Paris
   \bibitem[1986]{Mermilliod} Mermilliod, J.C. 1986,
      A\&AS, 63, 293
   \bibitem[1998]{Metzger et al.} Metzger, M.R., Caldwell, J.A.R., \&
                                  Schechter, P.L. 1998,
      AJ, 115, 635
   \bibitem[1997]{Mishurov et al.1} Mishurov, Yu.N., Zenina, I.A., Dambis,
                                    A.K., Mel'nik, A.M., \& Rastorguev,
                                    A.S. 1997,
      A\&A, 323, 775
   \bibitem[1999]{Mishurov et al.2} Mishurov, Yu.N., \& Zenina, I.A. 1999,
      A\&A, 341, 81
   \bibitem[2001]{Olano} Olano, C.A. 2001,
      AJ, 121, 295
   \bibitem[1998]{Olling et al.} Olling, R.P., \& Merrifield, M.R. 1998,   
      MNRAS, 297, 943
   \bibitem[1994]{Pont et al.1} Pont, F., Mayor, M., \& Burki, G. 1994,
      A\&A, 285, 415
   \bibitem[1997]{Pont et al.2} Pont, F., Queloz, D., Bratschi, P., \&
                                Mayor, M. 1997,
      A\&A, 318, 416
   \bibitem[1992]{Press et al.} Press, W.H., Teukolsky, S.A., Vetterling,
                                W.T., \& Flannery, B.P. 1992,
      Numerical Recipes, Cambridge University Press, Cambridge
   \bibitem[1989]{Racine et al.} Racine, R., \& Harris, W.E. 1989,
      AJ, 98, 1609
   \bibitem[2001]{Rastorguev et al.} Rastorguev, A.S., Glushkova, E.V.,
                                     Zabolotskikh, M.V., \& Baumgardt,
                                     H. 2001,
      Astronomical and Astrophysical Transactions, in press
   \bibitem[1993]{Reid} Reid, M.J. 1993,
      ARA\&A, 31, 345
   \bibitem[1969]{Roberts1} Roberts Jr., W.W. 1969,
      ApJ, 158, 123
   \bibitem[1970]{Roberts2} Roberts Jr., W.W. 1970,
      The spiral structure of our galaxy, eds. W. Becker, G. Contopoulos,
      IAU Symposium 38, 415
   \bibitem[1977]{Rohlfs} Rohlfs, K. 1977,
      Lectures in density waves, Springer-Verlag, Berlin
   \bibitem[1997]{Royer} Royer, F. 1999,
      PhD Thesis, Observatoire de Paris-Meudon, France
   \bibitem[1967]{Schild} Schild R. 1967,
      ApJ, 148, 449
   \bibitem[1965]{Schmidt} Schmidt, M. 1965,
      Galactic structure, eds. A. Blaauw and M. Schmidt, University
      Chicago Press, Chicago
   \bibitem[1975]{Schmidt-Kaler} Schmidt-Kaler, T. 1975,
      Vistas Astron., 19, 69
   \bibitem[1956]{Stibbs} Stibbs, D.W.N. 1956,
      MNRAS, 116, 453
   \bibitem[1969]{Toomre} Toomre, A. 1969,
      ApJ, 158, 899
   \bibitem[2000]{Torra et al.} Torra, J., Fern\'andez, D., \& Figueras,
                                F. 2000,
      A\&A, 359, 82 (Paper I)
   \bibitem[1995]{Vallee} Vall\'ee, J.P. 1995,
      ApJ, 454, 119
   \bibitem[1969]{Yuan} Yuan, C. 1969,
      ApJ, 158, 889
\end{thebibliography}
\end{document}